\documentclass[aps,prx,reprint,superscriptaddress]{revtex4-2}

\usepackage{amsmath,amsthm,amssymb,graphicx,xcolor,float,makecell,multirow}
\usepackage{esint}
\usepackage[unicode=true]{hyperref}
\hypersetup{
     colorlinks=true,
     linkcolor=blue,
     urlcolor=blue,
 }
\usepackage{tikz-cd}
\usepackage{tikz}
\usetikzlibrary{shapes.geometric, arrows.meta, positioning}
\usepackage{mathrsfs,mathtools,mathdots,dsfont}
\usepackage{listings}
\usepackage{subfigure}

\begin{document}

\title{Systematic Extraction of Exact Yang-Mills Solutions via Algebraic Tensor Ring Decomposition}

\author{Yu-Xuan Zhang}
\affiliation{School of Physics, Nankai University, Tianjin 300071, People's Republic of China}

\author{Jing-Ling Chen}
\email{chenjl@nankai.edu.cn}
\affiliation{Theoretical Physics Division, Chern Institute of Mathematics, Nankai University, Tianjin 300071, People's Republic of China}

\date{\today}

\begin{abstract}
The non-linear nature of Yang-Mills theory presents a challenge for extracting exact classical solutions, which are useful for understanding non-perturbative vacuum structures. In this paper, an algebraic tensor ring decomposition framework is introduced to systematically map the non-linear partial differential equations (PDEs) of Yang-Mills theory into tractable differential-algebraic systems. By promoting static pure-gauge backgrounds to dynamical variables, the reference state acts as a geometric template whose Maurer-Cartan forms generate the algebraic cross-terms necessary to stabilize non-linear self-interactions. To analytically resolve the resulting differential ideals, specific differential-algebraic quotient rings are employed as evaluation tools, and the solution space is organized by an algebraic bifurcation analysis. Applying this framework, three distinct classes of exact solutions are extracted: (i) relativistic $SU(2)$ color waves evaluated over an elliptic quotient ring, where the differential ideal bifurcates into a Decoupled Branch and two Coupled Branches, the latter exhibiting mass gap generation; (ii) dynamical dyonic flux tubes obtained from a time-dependent helical template, where the Gauss law ideal bifurcates the system into Coulomb, Dyonic, and symmetric Meissner branches. In the Meissner branch, an Artinian asymptotic truncation yields Bessel-type exponential screening, stabilized by a temporal dominance condition; and (iii) dynamical $SU(3)$ configurations where the Gauss law ideal bifurcates the solution space into four distinct phases. The non-trivial branches enforce a kinetic cancellation mechanism that maps the amplitude dynamics onto a generalized $x^2y^2$ chaotic oscillator. Across these settings, the framework provides a methodical approach to characterize the classical solution space of strongly coupled gauge theories.
\end{abstract}

\maketitle

\section{Introduction}
\label{sec:intro}

The classical configuration space of non-Abelian gauge theories, such as Yang-Mills theory and Quantum Chromodynamics (QCD), possesses a complex geometric and topological structure. Standard perturbative quantization relies on expansions around the trivial vacuum ($A_\mu = 0$). While this framework is successful in the high-energy, asymptotically free regime, it captures only local quantum fluctuations and does not fully account for macroscopic, non-perturbative phenomena such as color confinement and dynamical mass generation. 

Historically, the discovery of exact classical solutions—such as the 't Hooft-Polyakov monopole~\cite{tHooft1974, Polyakov1974}, Wu-Yang monopole~\cite{WuYang1975}, BPST instantons~\cite{BPST1975}, and multipseudoparticles~\cite{Witten1977, Jackiw1977, Yang1977}—established that the classical Yang-Mills vacuum supports coherent topological states~\cite{Actor1979}. A systematic mapping of these classical configurations is a useful prerequisite for non-perturbative quantization. In the strongly coupled regime, the path integral is heavily influenced by extended classical saddle points and topological background fields~\cite{Seiberg1994, Kondo2001, Cho2002}. Therefore, characterizing exact classical solutions provides a basis for analyzing physical phenomena like mass gaps and color confinement that are not accessible through standard perturbation theory.

Extracting exact solutions from the coupled, non-linear partial differential equations (PDEs) of Yang-Mills theory requires specific mathematical treatments. The traditional methodology typically relies on ansätze based on maximal space-time symmetries~\cite{Forgacs1980} (e.g., spherical symmetry, cylindrical symmetry) to reduce the PDEs to ordinary differential equations (ODEs). A notable advancement in this direction was the ADHM construction~\cite{ADHM1978}, which reduced the self-dual Yang-Mills equations to linear algebra. Moreover, theoretical frameworks like the Cho-Faddeev-Niemi (CFN) decomposition~\cite{Cho1980, FaddeevNiemi1999} have isolated topological degrees of freedom by separating the gauge field into Abelian/topological backgrounds and valence gluons. However, despite these successes, geometric reduction often over-constrains the general dynamical system, precluding the existence of broader asymmetrically coupled configurations. 

In this work, we propose an algebraic ring decomposition framework—inspired by differential algebraic geometry—to investigate the exact solution space of Yang-Mills theory. Rather than treating the formulation of ansätze merely as a geometric symmetry-reduction technique, the physical state is structurally decomposed into three distinct mathematical domains: a parameter ring $\mathcal{A}$ accommodating the dynamical variables, an analytical base space $\mathcal{U}$ capturing the continuous spacetime dependencies, and an algebraic structure module $\mathfrak{g}$ representing the internal Lie algebra. By projecting the local spacetime calculus onto linearly independent bases, the highly nonlinear PDEs are systematically mapped into a set of differential-algebraic equations (DAEs). 

To systematically solve these differential ideals, we introduce the technique of evaluating the system over specific differential-algebraic quotient rings. Different rings serve distinct analytical purposes: for instance, elliptic rings are utilized to extract exact periodic wave structures, while Artinian rings (e.g., rings with nilpotent elements) provide an algebraic foundation for exact perturbative truncation and asymptotic analysis. Furthermore, to resolve the issue of overdetermination—where generalized ansätze yield ideals with no non-trivial roots—a dynamical deformation mechanism is introduced. By assigning variables to pure-gauge backgrounds, these reference states cease to be trivial and instead become dynamically active geometric templates. Their Maurer-Cartan forms participate in the non-Abelian commutators, generating the geometric cross-terms required to compensate for missing self-interactions.

The paper is organized as follows. In Section~\ref{sec:method}, the formal tensor state space and the mapping of differential dynamics to quotient rings are defined. In Section~\ref{sec:color_waves}, exact $SU(2)$ color waves are extracted via an elliptic quotient ring, and the differential ideal is shown to bifurcate into a Decoupled Branch and two Coupled Branches revealing mass gap generation. In Section~\ref{sec:static_flux_tubes}, a dynamical helical topological template is introduced to evade Derrick's theorem. The corresponding Gauss law ideal performs a topological bifurcation, and the symmetric Meissner branch is analyzed via Artinian truncation, providing a classical realization of the dyonic dual-superconductor picture. In Section~\ref{sec:su3_dynamics}, the paradigm is extended to $SU(3)$ theory, demonstrating how activating a spatial Cartan template produces a four-branch algebraic bifurcation and maps the dynamics onto the $x^2y^2$ chaotic oscillator. Finally, in Section~\ref{sec:discussion}, the phenomenological implications are discussed, demonstrating how these configurations dynamically generate mass gaps and structurally resist the Savvidy vacuum instability.

\section{Algebraic Tensor Mapping and Quotient Ring Evaluation}
\label{sec:method}

\subsection{Mapping Differential Dynamics to Differential Ideals}

To dissect the non-linear dynamics, the gauge field is decoupled into a formal tensor product space $\mathcal{V}_{\text{form}} = \mathcal{A} \otimes_{\mathbb{K}} \mathcal{U} \otimes_{\mathbb{K}} \mathfrak{g}$. Here, $\mathbb{K}$ denotes an algebraically closed ground field (e.g., $\mathbb{C}$ or $\mathbb{R}$ for physical fields). The space $\mathcal{A}$ is a commutative parameter ring generated by undetermined spacetime functions; $\mathcal{U}$ is spanned by analytical base functions $\{f_\alpha(x)\}$ capturing the continuous spacetime dependencies; and $\mathfrak{g}$ is spanned by the Lie algebra generators $\{T_K\}$.

A generic gauge field configuration $\Psi$ is formulated within this space. A spacetime derivative $\partial_\mu$ maps the space into itself via the Leibniz rule: $\partial_\mu(a \otimes f \otimes T) = \partial_\mu a \otimes f \otimes T + a \otimes \partial_\mu f \otimes T$. When the Yang-Mills operator $\mathcal{O}_I(A) = D^\mu F_{\mu\nu}$ acts on this generalized state, the result is expanded over the linearly independent bases of $\mathcal{U}$ and $\mathfrak{g}$:
\begin{equation}
\mathcal{O}_I(\Psi) = \sum_K \left( \sum_\alpha P_{I\alpha K} \otimes f_\alpha \right) \otimes T_K = 0,
\end{equation}
where $P_{I\alpha K}$ are differential polynomials containing the variables from $\mathcal{A}$ and their derivatives. The validity of the equations of motion over the continuous spacetime demands the identical vanishing of all coefficients, $P_{I\alpha K} = 0$. This projects the coupled PDEs into a differential ideal $\mathcal{I}_{\text{diff}} = \langle P_{I\alpha K} \rangle$.

While one can analyze these DAEs directly, an algebraic approach inspired by differential Galois theory is to evaluate the ideal over specific quotient rings $\mathcal{A} / \langle \mathcal{I}_{\text{rule}} \rangle$. By imposing a derivation rule (e.g., an elliptic curve equation or an Artinian truncation condition), the differential operators act as purely algebraic derivations within the ring. This essentially translates a complex differential integrability problem into a systematic algebraic coefficient-matching problem over polynomial rings, analogous to how the ADHM construction algebraically maps the self-dual sectors.

\subsection{Dynamical Deformation of Geometric Templates}

A generalized parameterization often yields an overdetermined system with no non-trivial roots ($1 \in \mathcal{I}_{\text{diff}}$). To inject the necessary non-linearity, background deformations are utilized. 

Consider a trivial or decoupled field $A_\mu^{\text{simple}}$. Upon applying a spacetime-dependent gauge transformation $U(x)$, the field transforms as $A_\mu^U = U^{-1} A_\mu^{\text{simple}} U + U^{-1} \partial_\mu U$. The Maurer-Cartan form $A_\mu^{\text{vac}} = U^{-1} \partial_\mu U$ represents a pure-gauge background. By assigning sets of variables $\{a_i(x)\}$ and $\{b_j(x)\}$ from the parameter ring $\mathcal{A}$ to both the transformed field components and the pure-gauge terms, the generalized ansatz is constructed as:
\begin{equation}
A_\mu(\mathcal{A}) = \sum_i a_i(x) \left( U^{-1} A_{\mu, i}^{\text{simple}} U \right) + \sum_j b_j(x) A_{\mu, j}^{\text{vac}}.
\end{equation}

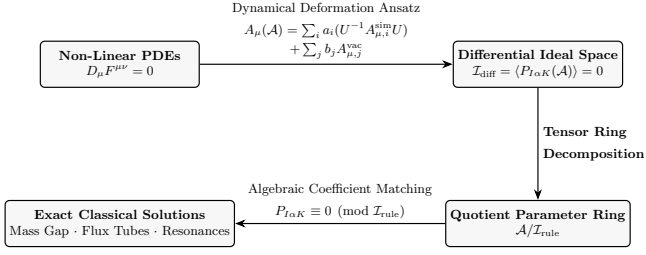
\begin{figure}[H]
    \centering
    \resizebox{1.0\linewidth}{!}{
    \begin{tikzpicture}[
        box/.style={rectangle, draw=black, thick, rounded corners=3pt, fill=black!3, align=center, minimum width=3.8cm, minimum height=1.1cm},
        arrow/.style={-{Stealth[scale=1.1]}, thick}
    ]
    
    \node[box] (pde) at (0, 0) {\textbf{Non-Linear PDEs}\\$D_\mu F^{\mu\nu} = 0$};
    
    \node[box] (ideal) at (10.0, 0) {\textbf{Differential Ideal Space}\\$\mathcal{I}_{\text{diff}} = \langle P_{I\alpha K}(\mathcal{A}) \rangle = 0$};
    
    \node[box] (ring) at (10.0, -3.8) {\textbf{Quotient Parameter Ring}\\$\mathcal{A}/\mathcal{I}_{\text{rule}}$};
    
    \node[box] (exact) at (0, -3.8) {\textbf{Exact Classical Solutions}\\Mass Gap $\cdot$ Flux Tubes $\cdot$ Resonances};
    
    \draw[arrow] (pde) -- node[above, align=center] {\small Dynamical Deformation Ansatz\\[1.5mm] \small $A_\mu(\mathcal{A}) = \sum_i a_i(U^{-1} A_{\mu,i}^{\text{sim}} U)$\\[1.0mm] \small $+ \sum_j b_j A_{\mu,j}^{\text{vac}}$} (ideal);
    
    \draw[arrow] (ideal) -- node[right, align=left] {\textbf{Tensor Ring}\\[1.5mm] \textbf{Decomposition}} (ring);
    
    \draw[arrow] (ring) -- node[above, align=center] {\small Algebraic Coefficient Matching\\[1.5mm] \small $P_{I\alpha K} \equiv 0 \pmod{\mathcal{I}_{\text{rule}}}$} (exact);
    
    \end{tikzpicture}
    }
    \caption{Workflow of the algebraic tensor ring decomposition framework. The non-linear Yang-Mills PDEs are parameterized over background geometric templates (dynamical deformation), projecting the coupled PDEs into a differential ideal space. Imposing specific algebraic quotient ideals transforms the differential equations into algebraic matching conditions, facilitating the extraction of exact classical solutions.}
    \label{fig:methodology}
\end{figure}

Elevating the constants $b_j$ to variables $b_j(x)$ implies that the term $\sum_j b_j(x) A_{\mu, j}^{\text{vac}}$ ceases to be pure gauge. Instead, the Maurer-Cartan form $A_{\mu, j}^{\text{vac}}$ serves as a geometric template. When evaluating the field strength $F_{\mu\nu} = \partial_\mu A_\nu - \partial_\nu A_\mu + [A_\mu, A_\nu]$, the non-vanishing Maurer-Cartan forms participate in the commutators. These commutators generate the geometric cross-terms required to compensate for the missing self-interaction potentials, thereby stabilizing the linear components into a non-linear dynamical state. The complete workflow of this algebraic tensor ring decomposition framework is illustrated in Fig.~\ref{fig:methodology}.

\section{Exact Relativistic Color Waves and Mass Gap Generation}
\label{sec:color_waves}

To demonstrate the application of the algebraic ring paradigm, exact traveling wave solutions in $SU(2)$ Yang-Mills theory are extracted. We adopt the Minkowski metric $\eta_{\mu\nu} = \text{diag}(1, -1, -1, -1)$ and anti-hermitian generators satisfying $[T_a, T_b] = \epsilon_{abc} T_c$.

\subsection{Ansatz Construction and Field Strength Evaluation}

A reference pure-gauge configuration dependent on a traveling phase $v = kz - \omega t$ is defined via the group element $U = \exp(v T_3)$. The generated background connection $A_\mu^{\text{vac}} = U^{-1} \partial_\mu U$ provides a longitudinal gauge skeleton: $A_t^{\text{vac}} = -\omega T_3$ and $A_z^{\text{vac}} = k T_3$. Variables $c_0(v), c_3(v)$ are assigned to the longitudinal skeleton, and $c_1(v), c_2(v)$ to the transverse directions aligned with the locally rotated bases $T_{1,2}(v) = U^{-1} T_{1,2} U$:
\begin{align}
A_t &= c_0(v) T_3, \quad A_z = c_3(v) T_3, \label{eq:ansatz_long}\\
A_x &= c_1(v) T_1(v), \quad A_y = c_2(v) T_2(v). \label{eq:ansatz_trans}
\end{align}
The longitudinal deformations from the pure-gauge vacuum are defined as $\Omega(v) = c_0(v) + \omega$ and $K(v) = c_3(v) - k$. 

The non-vanishing components of the color-electric field $F_{0i}$ and color-magnetic field $F_{ij}$ are explicitly calculated as:
\begin{align}
F_{0x} &= \omega c_1' T_1(v) - \Omega c_1 T_2(v), \\
F_{0y} &= \omega c_2' T_2(v) + \Omega c_2 T_1(v), \\
F_{0z} &= (\omega K' + k \Omega') T_3, \\
F_{yz} &= k c_1' T_1(v) + K c_1 T_2(v), \\
F_{zx} &= k c_2' T_2(v) - K c_2 T_1(v), \\
F_{xy} &= c_1 c_2 T_3.
\end{align}
Here, the longitudinal electric field $F_{0z}$ is generated by the gradients of the longitudinal deformations. In contrast, the transverse fields contain both derivative terms and algebraic cross-terms (e.g., $\Omega c_1 T_2(v)$) originating from the commutators between the longitudinal background and the transverse wave.

\subsection{Differential Ideals and Elliptic Ring Evaluation}

Substituting these fields into the Yang-Mills equations $D_\mu F^{\mu\nu} = 0$, the spacetime derivatives of the rotated bases $\partial_\mu T_{1,2}(v)$ are algebraically absorbed by the commutators with $A_\mu^{\text{vac}}$. The $\nu = t$ and $\nu = z$ components generate the longitudinal differential ideals:
\begin{align}
\mathcal{I}_t&: \quad k(\omega c_3'' + k c_0'') - \Omega (c_1^2 + c_2^2) = 0, \label{eq:It}\\
\mathcal{I}_z&: \quad \omega(\omega c_3'' + k c_0'') + K (c_1^2 + c_2^2) = 0. \label{eq:Iz}
\end{align}
Eliminating the second-derivative terms yields the algebraic constraint $(\omega \Omega + k K)(c_1^2 + c_2^2) = 0$. For any non-trivial wave ($c_1^2 + c_2^2 \neq 0$), this enforces the kinematic relation $K(v) = -\frac{\omega}{k}\Omega(v)$. Applying this relation to the transverse ($\nu = x, y$) components maps the dynamics to a coupled non-linear oscillator system:
\begin{align}
(\omega^2 - k^2) \Omega'' + \Omega (c_1^2 + c_2^2) &= 0, \label{eq:sys_omega}\\
(\omega^2 - k^2) c_1'' + c_1 (c_2^2 + \gamma \Omega^2) &= 0, \label{eq:sys_c1}\\
(\omega^2 - k^2) c_2'' + c_2 (c_1^2 + \gamma \Omega^2) &= 0, \label{eq:sys_c2}
\end{align}
where $\gamma = (\omega^2 - k^2)/k^2$ acts as a geometric coupling constant. 

To extract exact periodic structures, we evaluate this system over an elliptic quotient ring $\mathcal{R}_{\text{ellip}} = \mathbb{R}[f, f'] / \langle (f')^2 + \frac{\lambda}{2} f^4 - E \rangle$, where the derivation rule is defined as $f'' = -\lambda f^3$. By constraining the variables to proportional rays $c_1(v) = f(v)$, $c_2(v) = \kappa f(v)$, and $\Omega(v) = \alpha f(v)$, the differential equations reduce to matching the cubic coefficients in $\mathcal{R}_{\text{ellip}}$.

\subsection{Exact Solutions and Physical Analysis}

The algebraic consistency of the cubic coefficients bifurcates the solution space. To analyze the physical properties, we evaluate the gauge-invariant energy-momentum tensor $T_{\mu\nu} = 2 \text{tr}(F_{\mu\lambda} F_{\nu}^{\ \lambda} - \frac{1}{4}\eta_{\mu\nu} F_{\alpha\beta} F^{\alpha\beta})$. 

Since all non-trivial branches are governed by the differential equation $f'' = -\lambda f^3$, the system possesses a conserved oscillator energy defined by the first integral:
\begin{equation}
E_0 = (f')^2 + \frac{\lambda}{2} f^4.
\end{equation}
This constant $E_0$ represents the intrinsic excitation intensity of the wave. We utilize $E_0$ to systematically express the macroscopic energy density $\rho = T_{00}$ and momentum flux $S_z = T_{0z}$ for each branch.

\subsubsection{Branch A: Pure Transverse Self-Interacting Wave ($\Omega = 0$)}

Setting $\alpha = 0$ yields $\Omega(v) = K(v) = 0$. Consistency requires $\kappa = \pm 1$ and $\lambda = 1/(\omega^2 - k^2)$. The exact solution corresponds to the Jacobi elliptic function $f(v) = A \, \text{cn}(\lambda^{1/2} A v, 1/\sqrt{2})$. Substituting $(f')^2 = E_0 - \frac{1}{2(\omega^2 - k^2)} f^4$ into the tensor components yields:
\begin{align}
\rho_{\text{A}} &= (\omega^2 + k^2) E_0 - \frac{k^2}{\omega^2 - k^2} f^4, \\
S_{z, \text{A}} &= 2 \omega k E_0 - \frac{\omega k}{\omega^2 - k^2} f^4.
\end{align}
This branch represents a standard transverse non-Abelian plane wave. The requirement of a non-trivial solution enforces a timelike dispersion $\omega^2 > k^2$, indicating that an effective mass ($m_{\text{eff}}^2 = \omega^2 - k^2 > 0$) is generated by the transverse magnetic self-interaction. The negative sign preceding the $f^4$ term indicates that the macroscopic energy density oscillates out of phase with the field amplitude, characterizing a kinetically dominated wave.

\subsubsection{Branch B: Isotropic Cartan-Coupled Wave ($\Omega \neq 0$, $\kappa = \pm 1$)}

Assuming $\alpha \neq 0$ and $\kappa = \pm 1$, consistency demands $\alpha^2 = k^2/(\omega^2 - k^2)$ and $\lambda = 2/(\omega^2 - k^2)$. Applying the corresponding substitution $(f')^2 = E_0 - \frac{1}{\omega^2 - k^2} f^4$, the $f^4$ terms cancel out:
\begin{align}
\rho_{\text{B}} &= \frac{1}{2}(3\omega^2 + k^2) E_0, \\
S_{z, \text{B}} &= 2 \omega k E_0.
\end{align}
In this isotropically polarized branch, the transverse wave drives the longitudinal gauge skeleton $\Omega(v)$ into synchronized oscillation. The cancellation of the $f^4$ terms demonstrates that the kinetic and potential energy fluctuations maintain a dynamic balance. Consequently, the macroscopic energy density and momentum flux are constant in spacetime, analogous to the behavior of a circularly polarized wave.

\subsubsection{Branch C: Linearly Polarized Cartan-Induced Wave ($\kappa = 0$)}

Setting $\kappa = 0$ turns off one transverse polarization ($c_2 = 0$). Consistency yields $\lambda = 1/(\omega^2 - k^2)$ and $\gamma \alpha^2 = 1$. The densities evaluate to:
\begin{align}
\rho_{\text{C}} &= \omega^2 E_0 + \frac{k^2}{2(\omega^2 - k^2)} f^4, \\
S_{z, \text{C}} &= \omega k E_0 + \frac{\omega k}{2(\omega^2 - k^2)} f^4.
\end{align}
In standard Yang-Mills theory, a single-direction transverse field yields a trivial Abelian dispersion. Here, the dynamical deformation $\Omega(v) \neq 0$ provides an induced potential that sustains the linearly polarized wave and endows it with mass. The positive sign preceding the $f^4$ term indicates that the energy density oscillates in phase with the field amplitude, characterizing a potential-energy dominated wave.

\subsubsection{Energy-Momentum Sum Rule}

\begin{figure}[H]
    \centering
    \includegraphics[width=0.9\linewidth]{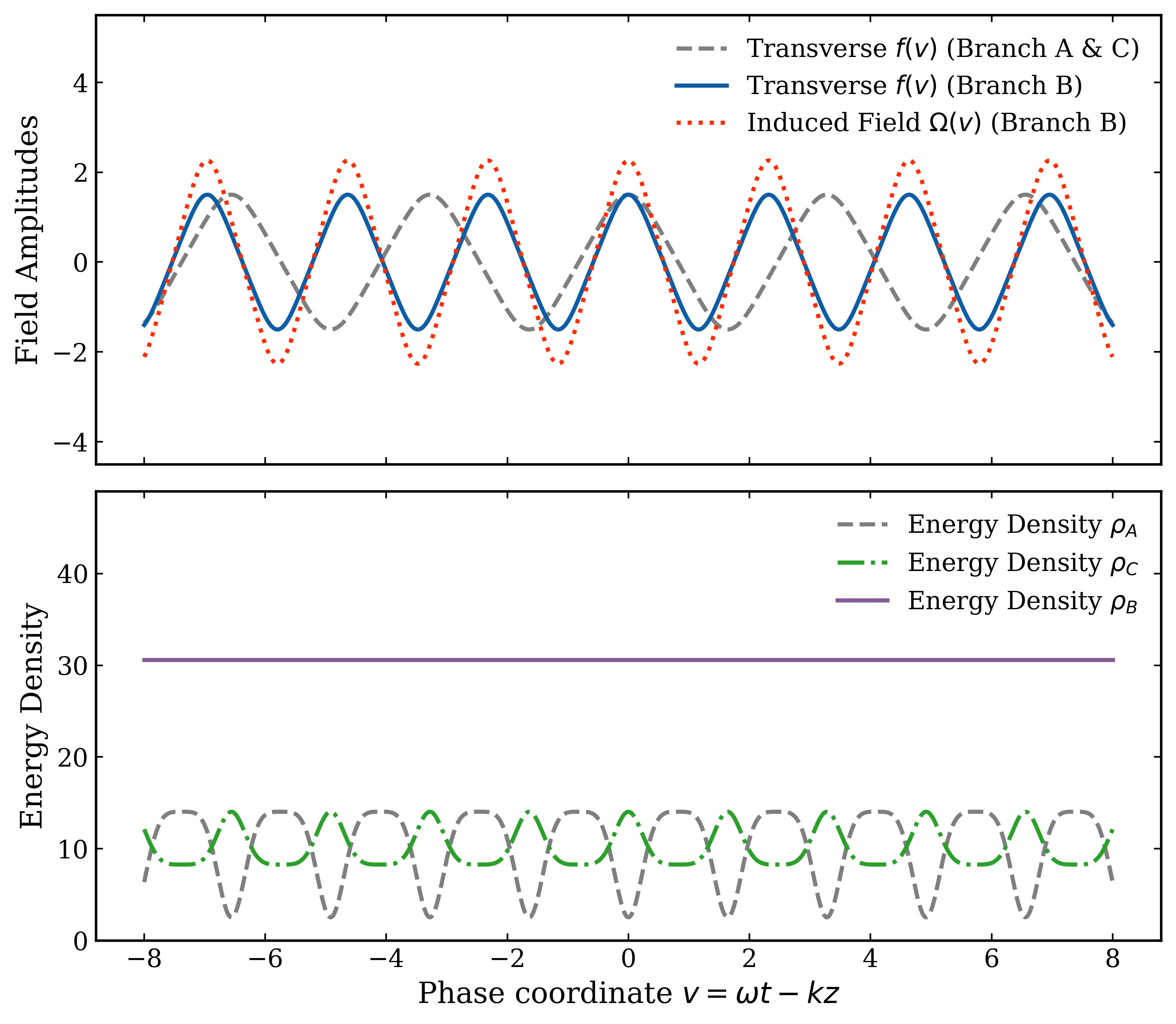}
    \caption{Energy density and field amplitudes of the exact $SU(2)$ color waves. (Top panel) The transverse wave amplitudes $f(v)$ for Branch A and C are mathematically identical, while Branch B exhibits a modified period and is accompanied by a dynamically induced longitudinal field $\Omega(v)$. (Bottom panel) The coupled energy densities exhibit non-zero baselines, demonstrating the dynamical generation of a mass gap. The isotropic coupling in Branch B balances the kinetic and potential energy fluctuations, rendering its total energy density constant. Meanwhile, the kinetically dominated Branch A and the potential-dominated Branch C exhibit out-of-phase oscillating energy densities, which collectively satisfy the sum rule $\frac{1}{2}\rho_{\text{A}} + \rho_{\text{C}} = \rho_{\text{B}}$.}
    \label{fig:color_wave}
\end{figure}

Comparing the macroscopic densities of the three branches reveals a linear relation. Despite the non-linear nature of the underlying Yang-Mills equations, the energy densities and momentum fluxes satisfy the superposition rule:
\begin{equation}
\frac{1}{2}\rho_{\text{A}} + \rho_{\text{C}} = \rho_{\text{B}}, \quad \frac{1}{2}S_{z, \text{A}} + S_{z, \text{C}} = S_{z, \text{B}}.
\end{equation}
This sum rule indicates that, at the level of the macroscopic energy-momentum tensor, the isotropically coupled state (Branch B) is mathematically equivalent to the linear combination of half a pure transverse wave (Branch A) and a linearly polarized induced wave (Branch C). The field amplitudes and the exact energy superposition of these branches are depicted in Fig.~\ref{fig:color_wave}.

\section{Dynamical Helical Configurations and Topological Bifurcation}
\label{sec:static_flux_tubes}

In 3D spatial pure Yang-Mills theory, Derrick's theorem precludes the existence of finite-energy, purely static topological solitons. Spatial scaling perturbations drive static non-Abelian configurations toward collapse or dispersion. To evade this scale instability without introducing external scalar fields, dynamical time-dependence can be injected into the geometric skeleton. In this section, the algebraic ring decomposition is applied in cylindrical coordinates $(t, r, \phi, z)$ using a dynamical helical template, demonstrating how a temporal phase stabilizes the topological flux tube and induces algebraic bifurcations.

\subsection{Dynamical Ansatz Construction and Field Strength Evaluation}

To construct a stationary but non-static configuration, a time-dependent helical pure-gauge background is introduced via the gauge transformation $U = \exp[(n\phi + kz - \omega t)T_3]$, where $n$ is the azimuthal winding number, $k$ is the longitudinal wave number, and $\omega$ is the temporal frequency. The Maurer-Cartan forms $A_\mu^{\text{vac}} = U^{-1}\partial_\mu U$ provide the spacetime skeleton:
\begin{equation}
A_t^{\text{vac}} = -\omega T_3, \quad A_\phi^{\text{vac}} = n T_3, \quad A_z^{\text{vac}} = k T_3.
\end{equation}

Following the algebraic mapping paradigm, this skeleton is dynamically activated by assigning radial functions $K(r)$, $\Phi(r)$, and $W(r)$ to the longitudinal components. To incorporate valence gluons, transverse magnetic condensates $c_1(r)$ and $c_2(r)$ are introduced, aligned with the locally rotated bases $T_{1,2}(t, \phi, z) = U^{-1} T_{1,2} U$. The generalized ansatz is constructed in the original frame as:
\begin{align}
A_t &= K(r) T_3, \label{eq:dyn_At} \\
A_\phi &= \Phi(r) T_3 + r c_1(r) T_1(t, \phi, z), \label{eq:dyn_Aphi} \\
A_z &= W(r) T_3 + c_2(r) T_1(t, \phi, z). \label{eq:dyn_Az}
\end{align}

The spacetime derivatives of the rotated bases generate the geometric cross-terms: $\partial_t T_1 = \omega T_2$, $\partial_\phi T_1 = -n T_2$, and $\partial_z T_1 = -k T_2$. Defining the longitudinal deformations from the vacuum skeleton as $\Omega(r) = K(r) + \omega$, $\tilde{\Phi}(r) = \Phi(r) - n$, and $\tilde{W}(r) = W(r) - k$, the non-vanishing field strength components $F_{\mu\nu} = \partial_\mu A_\nu - \partial_\nu A_\mu + [A_\mu, A_\nu]$ evaluate to:
\begin{align}
F_{tr} &= \Omega' T_3, \quad F_{t\phi} = \Omega r c_1 T_2, \quad F_{tz} = \Omega c_2 T_2, \\
F_{r\phi} &= \tilde{\Phi}' T_3 + (r c_1)' T_1, \quad F_{rz} = \tilde{W}' T_3 + c_2' T_1, \\
F_{\phi z} &= (\tilde{\Phi} c_2 - r c_1 \tilde{W}) T_2.
\end{align}

\subsection{Differential Ideals and Exhaustive Bifurcation}

Substituting these field strengths into the Yang-Mills equations $D_\mu F^{\mu\nu} = 0$, the coupled PDEs are projected into algebraic and differential ideals.

The temporal component ($\nu = t$, Gauss law) yields:
\begin{equation}
\left[ \Omega'' + \frac{1}{r}\Omega' + \Omega(c_1^2 + c_2^2) \right] T_3 - \frac{\Omega}{r} \left( \tilde{\Phi} c_1 + r \tilde{W} c_2 \right) T_1 = 0.
\end{equation}
The vanishing of the $T_1$ component yields a primary algebraic Gauss ideal:
\begin{equation}
\mathcal{I}_{\text{Gauss}} : \quad \Omega \left( \tilde{\Phi} c_1 + r \tilde{W} c_2 \right) = 0. \label{eq:dyn_Gauss_ideal}
\end{equation}

Evaluating the radial Amp\`{e}re's law ($D_\mu F^{\mu r} = 0$) and isolating the $T_2$ cross-terms reveals a secondary differential-algebraic ideal for the transverse condensates:
\begin{equation}
\mathcal{I}_{\text{Radial}} : \quad (r c_1)^2 \left( \frac{\tilde{\Phi}}{r c_1} \right)' + r^2 c_2^2 \left( \frac{\tilde{W}}{c_2} \right)' = 0. \label{eq:dyn_Radial_ideal}
\end{equation}

The simultaneous resolution of the primary ideal $\mathcal{I}_{\text{Gauss}}$ and the secondary ideal $\mathcal{I}_{\text{Radial}}$ bifurcates the solution space into distinct branches.

\subsection{Solutions and Phase Analysis}

\subsubsection{Branch A: Decoupled Coulomb Phase ($c_1 = c_2 = 0$)}

If the transverse condensates vanish, the non-linear cross-terms in the ideals disappear. The equations decouple into standard Laplace equations (e.g., $\Omega'' + \Omega'/r = 0$), yielding unconfined logarithmic potentials $\Omega \sim \ln r$. This represents a perturbative, non-interacting vacuum state.

\subsubsection{Branch B: Helical Dyonic Phase ($c_1 \neq 0, c_2 \neq 0, \Omega \neq 0$)}

Assuming a non-zero temporal deformation ($\Omega \neq 0$) and active transverse fields, $\mathcal{I}_{\text{Gauss}}$ algebraically demands the locking relation $\frac{\tilde{\Phi}}{r c_1} = - \frac{\tilde{W}}{c_2} \equiv \Lambda(r)$. Substituting this ratio into the secondary ideal $\mathcal{I}_{\text{Radial}}$ yields the constraint:
\begin{equation}
r^2 (c_1^2 - c_2^2) \Lambda' = 0.
\end{equation}
This condition leads the Dyonic phase to bifurcate into two sub-branches. For the isotropic transverse phase (Sub-branch B1), the condition is satisfied by $c_1 = \pm c_2$. This enforces the algebraic locking $\tilde{\Phi}(r) = \mp r \tilde{W}(r)$. In this state, the transverse magnetic condensates are isotropically distributed between the azimuthal and longitudinal planes, locking the two topological templates into a synchronized spatial wave.

Alternatively, for the proportional longitudinal phase (Sub-branch B2), the ideal is satisfied by $\Lambda' = 0$, implying $\Lambda$ is a constant. This yields the linear relations $\tilde{\Phi}(r) = \lambda r c_1(r)$ and $\tilde{W}(r) = -\lambda c_2(r)$. The longitudinal magnetic templates become proportional to their respective transverse condensates, linearizing the coupled non-Abelian interactions.

\subsubsection{Branch C: Symmetric Meissner Branch ($c_2 = 0, \Omega \neq 0$)}

Setting $c_2 = 0$ and $\Omega \neq 0$, the primary ideal $\mathcal{I}_{\text{Gauss}}$ enforces $\tilde{\Phi} = 0$ (implying $\Phi(r) = n$). The azimuthal magnetic field is screened and locked to the topological vacuum skeleton. Evaluating the remaining spatial Amp\`{e}re's laws under this constraint yields a symmetric set of coupled differential equations:
\begin{align}
\Omega'' + \frac{1}{r} \Omega' + c_1^2 \Omega &= 0, \label{eq:branchC_Omega} \\
\tilde{W}'' + \frac{1}{r} \tilde{W}' - c_1^2 \tilde{W} &= 0, \label{eq:branchC_W} \\
c_1'' + \frac{1}{r} c_1' - \frac{1}{r^2} c_1 - (\Omega^2 - \tilde{W}^2) c_1 &= 0. \label{eq:branchC_c1}
\end{align}
The temporal deformation $\Omega^2$ provides a positive effective mass squared that assists in evading Derrick's theorem, while $\tilde{W}^2$ contributes a tachyonic term. The physical properties are analyzed through the asymptotic behaviors:

\textbf{Core Region ($r \to 0$):} To maintain regularity against the $1/r^2$ centrifugal barrier in Eq.~(\ref{eq:branchC_c1}), the condensate scales as $c_1(r) \sim \beta r$. Substituting this into the longitudinal equations yields the Taylor expansion $\Omega(r) \simeq \Omega_0 - \frac{1}{4}\beta^2 \Omega_0 r^2$. The transverse condensate vanishes at the origin ($c_1 \to 0$), indicating that the non-Abelian gauge symmetry is locally restored at the center of the flux tube.

\textbf{Artinian Asymptotic Vacuum ($r \to \infty$):} Due to the opposing signs of the mass-squared terms ($+c_1^2$ versus $-\Omega^2$), the system lacks a global Bogomolny-Prasad-Sommerfield (BPS) reduction. Therefore, an Artinian quotient ring is utilized to extract the asymptotic tail behavior. Evaluating the system over an Artinian ring defined by the nilpotent ideal $\langle c_1^2 \rangle = 0$, Eqs.~(\ref{eq:branchC_Omega}) and (\ref{eq:branchC_W}) reduce to Laplace equations, requiring $\Omega(r) \to \Omega_\infty$ and $\tilde{W}(r) \to \tilde{W}_\infty$ to maintain finite energy. The condensate equation (\ref{eq:branchC_c1}) then truncates to a modified Bessel equation:
\begin{equation}
c_1'' + \frac{1}{r} c_1' - \left( M^2 + \frac{1}{r^2} \right) c_1 = 0, \quad M^2 \equiv \Omega_\infty^2 - \tilde{W}_\infty^2.
\end{equation}
For the solution to be localized, the system requires a temporal dominance condition $M^2 > 0$, meaning the temporal phase excitation exceeds the longitudinal magnetic field at spatial infinity. Under this condition, the solution in the Artinian ring is the modified Bessel function of the second kind:
\begin{equation}
c_1(r) \propto K_1(M r) \sim \sqrt{\frac{\pi}{2 M r}} e^{-M r}.
\end{equation}
In contrast to purely static configurations that yield power-law screening, the dynamical injection of the temporal phase $\omega$ introduces a positive effective mass $M$. The Meissner branch connects the symmetry-restored core to an asymptotic vacuum characterized by Bessel-type exponential screening.

\subsubsection{Branch D: Pure Magnetic Phase ($\Omega = 0$)}

If the temporal deformation vanishes ($\Omega(r) = K(r) + \omega = 0$), the longitudinal connection operates as a background gauge ($A_t = -\omega T_3$). The color-electric field $F_{ti}$ is extinguished. Consequently, the primary Gauss ideal $\mathcal{I}_{\text{Gauss}}$ (Eq.~\ref{eq:dyn_Gauss_ideal}) reduces to $0 = 0$, removing the constraint $\tilde{\Phi} c_1 + r \tilde{W} c_2 = 0$.

\begin{figure}[H]
    \centering
    \includegraphics[width=0.9\linewidth]{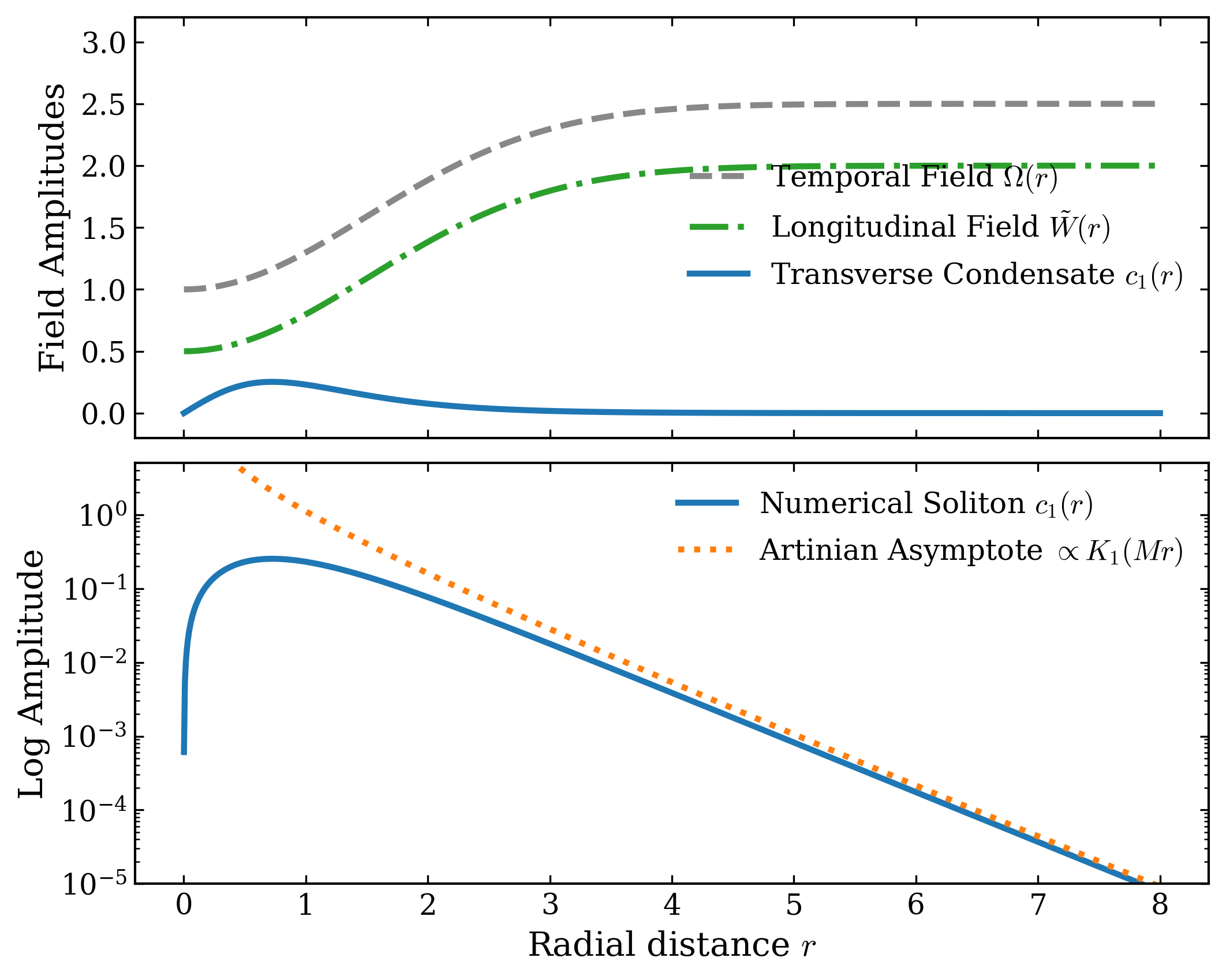}
    \caption{Radial profiles and asymptotic validation of the $SU(2)$ helical flux tube (Branch C). (Top panel) The longitudinal fields $\Omega(r)$, $\tilde{W}(r)$ and the transverse condensate $c_1(r)$ interpolate between the symmetry-restored core and the topological vacuum. At the center ($r \to 0$), the condensate vanishes as $c_1 \propto r$, effectively restoring the non-Abelian gauge symmetry. (Bottom panel) Log-amplitude comparison between the numerical soliton and the analytical Artinian asymptote. In the far-field regime ($r > 3$), the numerical solution (solid blue) aligns with the modified Bessel function $K_1(Mr)$ (dotted orange) derived from the Artinian quotient ring. This confirms that the temporal dominance condition $M^2 = \Omega_\infty^2 - \tilde{W}_\infty^2 > 0$ generates a mass gap, leading to the exponential screening of the color-magnetic flux and the stabilization of the soliton against Derrick's theorem.}
    \label{fig:meissner}
\end{figure}

Under symmetric spatial conditions ($c_2 = 0$, $\tilde{\Phi} = 0$), the remaining spatial Amp\`{e}re equations reduce to:
\begin{align}
\tilde{W}'' + \frac{1}{r} \tilde{W}' - c_1^2 \tilde{W} &= 0, \\
c_1'' + \frac{1}{r} c_1' - \frac{1}{r^2} c_1 + \tilde{W}^2 c_1 &= 0.
\end{align}
This coupled system takes a form analogous to the classical Abelian Higgs model evaluated at the Prasad-Sommerfield limit (BPS vortex equations). In this phase, the longitudinal magnetic skeleton $\tilde{W}$ acts analogously to a symmetry-breaking scalar field, while the transverse condensate $c_1$ acts as the gauge field, supporting a pure magnetic topological vortex without external scalar fields.
The radial profiles and the numerical validation of this Artinian asymptotic behavior are shown in Fig.~\ref{fig:meissner}.

\section{Dynamical Generation of Cartan Fields in $SU(3)$ Theory}
\label{sec:su3_dynamics}

To further demonstrate the versatility of the algebraic ring framework, we extend the paradigm to $SU(3)$ Yang-Mills theory. We show how the activation of a spatial Cartan template preserves the kinetic cancellation mechanism, dynamically mapping the coupled gauge system onto a generalized $x^2 y^2$ chaotic oscillator.

\subsection{Spatial Cartan Template and Field Strengths}

We utilize the anti-hermitian generators for $\mathfrak{su}(3)$, where $T_3$ and $T_8$ span the commuting Cartan subalgebra. The relevant commutation relations connecting the Cartan subalgebra to the V-spin and U-spin sectors are $[T_4, T_5] = \frac{1}{2} T_3 + \frac{\sqrt{3}}{2} T_8 \equiv Y_V$ and $[T_6, T_7] = -\frac{1}{2} T_3 + \frac{\sqrt{3}}{2} T_8 \equiv Y_U$.

We introduce a purely spatial pure-gauge background via $U = \exp(k_1 z T_3 + k_2 z T_8)$, whose Maurer-Cartan form $A_z^{\text{vac}} = k_1 T_3 + k_2 T_8$ acts as the geometric skeleton. We dynamically activate this skeleton by promoting the constants to time-dependent functions $K_1(t)$ and $K_2(t)$. A temporal Cartan deformation $A_t = c_0(t) T_3 + \Gamma(t) T_8$ and time-dependent transverse deformations in both V-spin and U-spin sectors are introduced:
\begin{align}
A_t &= c_0(t) T_3 + \Gamma(t) T_8, \\
A_x &= c_1(t) T_4 + c_2(t) T_5, \\
A_y &= c_3(t) T_6 + c_4(t) T_7, \\
A_z &= K_1(t) T_3 + K_2(t) T_8.
\end{align}
We define the effective temporal Cartan couplings as $\tilde{c}_V = \frac{1}{2} c_0 + \frac{\sqrt{3}}{2} \Gamma$ and $\tilde{c}_U = -\frac{1}{2} c_0 + \frac{\sqrt{3}}{2} \Gamma$, and parameterize the transverse fields in polar coordinates: $c_1 + i c_2 = R_V e^{i\theta_V}$ and $c_3 + ic_4 = R_U e^{i\theta_U}$. 

This parameterization triggers a \textit{kinetic cancellation mechanism}. When the algebraic locking conditions $\tilde{c}_V = -\dot{\theta}_V$ and $\tilde{c}_U = -\dot{\theta}_U$ hold, the non-Abelian commutators (e.g., $[A_t, A_x]$) cancel the temporal derivatives of the phases. Consequently, the coupled field strengths are stripped of all rotational components, reducing to purely radial oscillating expressions.

\subsection{Differential Ideals and Exhaustive Bifurcation}

Substituting these field strengths into the temporal Yang-Mills equation (Gauss law, $D_i F^{it} = 0$), the spatial derivatives vanish since the fields are purely time-dependent. The constraint projects entirely into a primary algebraic ideal:
\begin{equation}
\mathcal{I}_{\text{Gauss}}: \quad \langle R_V^2 (\dot{\theta}_V + \tilde{c}_V),\; R_U^2 (\dot{\theta}_U + \tilde{c}_U) \rangle = 0. \label{eq:su3_ideal}
\end{equation}
Analogous to the topological bifurcation in Section \ref{sec:static_flux_tubes}, this Gauss ideal bifurcates the $SU(3)$ dynamics into four distinct physical branches:

\subsubsection{Branch A: Trivial Cartan Phase ($R_V = R_U = 0$)}

The transverse fields vanish, and the non-linear cross-terms disappear. The activated Cartan templates $K_1(t), K_2(t)$ decouple, evolving as trivial free fields representing a non-interacting vacuum.

\subsubsection{Branch B: V-spin Resonance Phase ($R_V \neq 0, R_U = 0$)}

Assuming the U-spin sector is dormant, the ideal enforces the kinetic cancellation condition $\tilde{c}_V = -\dot{\theta}_V$. The internal phase rotation of the V-spin gluons dynamically sources the Cartan temporal field. 

Focusing on this branch, we define the effective spatial coupling $\tilde{K}_V = \frac{1}{2} K_1 + \frac{\sqrt{3}}{2} K_2$. The remaining spatial Yang-Mills equations evaluate to a symmetric non-linear coupled system:
\begin{align}
\ddot{R}_V + \tilde{K}_V^2 R_V &= 0, \label{eq:su3_R}\\
\ddot{\tilde{K}}_V + R_V^2 \tilde{K}_V &= 0. \label{eq:su3_K}
\end{align}
This constitutes the $x^2 y^2$ chaotic oscillator. The algebraic activation of the spatial Cartan template promotes the geometric background $\tilde{K}_V$ from a rigid constant to a dynamical field, which exchanges non-linear energy with the transverse amplitude $R_V(t)$.

To extract exact coherent states, we evaluate this system over the elliptic quotient ring $\mathcal{R}_{\text{ellip}} = \mathbb{R}[f, f'] / \langle (f')^2 + \frac{1}{2} f^4 - E_0 \rangle$, with the derivation rule $f'' = -f^3$. Constraining the variables to the symmetric ray $R_V = \tilde{K}_V = f(t)$, both differential equations map identically into the elliptic ideal, yielding the analytical solution $f(t) = A\,\text{cn}(At, 1/\sqrt{2})$.

The dynamics are governed by the conserved oscillator energy $E = \frac{1}{2}\dot{R}_V^2 + \frac{1}{2}\dot{\tilde{K}}_V^2 + \frac{1}{2}\tilde{K}_V^2 R_V^2 = \text{const}$. Furthermore, the kinetic cancellation condition implies the existence of a conserved macroscopic isospin charge:
\begin{equation}
Q = R_V^2 \dot{\theta}_V = \text{const}.
\end{equation}
The non-linearity of the Yang-Mills equations is absorbed by the free internal phase rotation $\theta_V(t)$, demonstrating a self-sustaining coherent resonance.

\subsubsection{Branch C: U-spin Resonance Phase ($R_V = 0, R_U \neq 0$)}

Symmetric to Branch B, the Gauss ideal enforces $\tilde{c}_U = -\dot{\theta}_U$. The U-spin transverse fields exchange energy with the corresponding spatial Cartan coupling $\tilde{K}_U = -\frac{1}{2} K_1 + \frac{\sqrt{3}}{2} K_2$, forming an identical $x^2y^2$ resonance.

\subsubsection{Branch D: Fully Coupled $SU(3)$ Phase ($R_V \neq 0, R_U \neq 0$)}

When both V-spin and U-spin sectors are actively excited, the Gauss ideal demands both conditions simultaneously. Solving for the Cartan temporal fields yields the algebraic locking: $c_0 = -(\dot{\theta}_V - \dot{\theta}_U)$ and $\Gamma = -\frac{1}{\sqrt{3}}(\dot{\theta}_V + \dot{\theta}_U)$. 

\begin{figure}[H]
    \centering
    \includegraphics[width=1.0\linewidth]{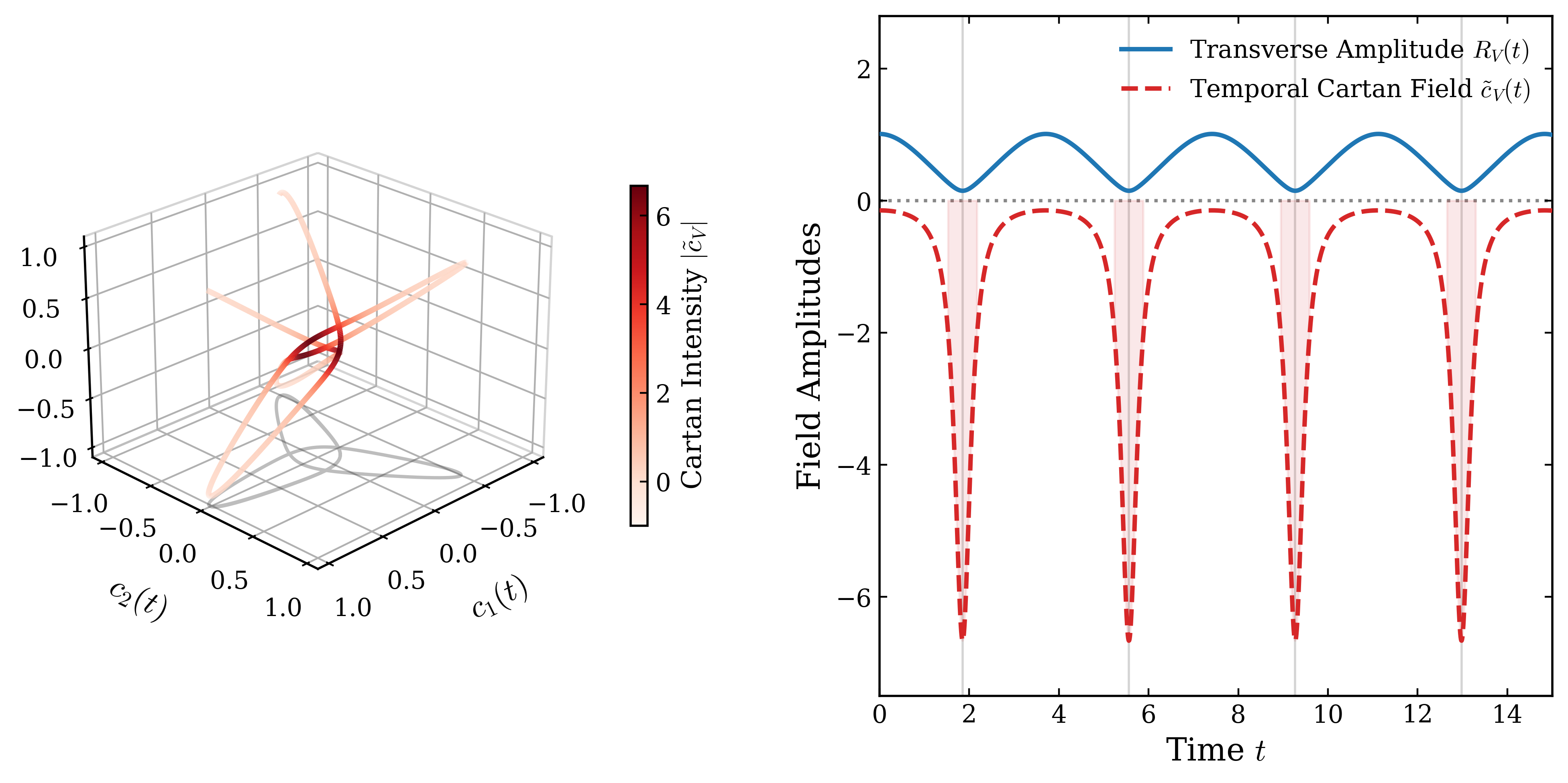}
    \caption{Dynamical trajectory and temporal evolution of the $SU(3)$ coherent resonance (Branch B). (Left panel) The 3D orbital trajectory parameterized by the transverse fields $(c_1, c_2)$ and the spatial Cartan coupling $\tilde{K}_V$. The bottom 2D shadow projection reveals the precessing petal-like structure of the generalized $x^2y^2$ chaotic attractor, decoupled from longitudinal fluctuations. The color gradient maps the absolute intensity of the induced temporal Cartan field $|\tilde{c}_V|$. (Right panel) Time evolution demonstrating the kinetic cancellation mechanism. Vertical guide lines connect the moments when the transverse amplitude $R_V(t)$ reaches its centrifugal minimum to the generation of sharp resonant spikes in the Cartan field $\tilde{c}_V(t)$. This physically illustrates a self-sustaining non-linear energy exchange within the topological $SU(3)$ vacuum.}
    \label{fig:su3_resonance}
\end{figure}

In this fully coupled phase, the spatial equations project into a generalized 4-component chaotic network:
\begin{align}
\ddot{R}_V + \tilde{K}_V^2 R_V &= 0, \\
\ddot{R}_U + \tilde{K}_U^2 R_U &= 0, \\
\ddot{K}_1 + \frac{1}{2} R_V^2 \tilde{K}_V - \frac{1}{2} R_U^2 \tilde{K}_U &= 0, \\
\ddot{K}_2 + \frac{\sqrt{3}}{2} R_V^2 \tilde{K}_V + \frac{\sqrt{3}}{2} R_U^2 \tilde{K}_U &= 0.
\end{align}
In this regime, the spatial Cartan fields $K_1(t)$ and $K_2(t)$ act as dynamical mediators. Although the V-spin ($R_V$) and U-spin ($R_U$) amplitudes do not interact directly, they are coupled through their mutual interaction with the shared Cartan skeleton. This demonstrates that the algebraic ring framework captures not only isolated coherent states but also the interlinked chaotic energy exchange native to $SU(3)$ theories. The 3D orbital trajectory and the temporal evolution of this kinetic cancellation mechanism are visualized in Fig.~\ref{fig:su3_resonance}.

\section{Discussion and Phenomenological Implications}
\label{sec:discussion}

The extraction of classical configurations via the algebraic quotient ring framework provides analytical insights into non-perturbative Yang-Mills dynamics. To evaluate the physical utility of this method, it is necessary to examine how these classical solutions connect to established phenomenological models and to verify their quantum structural stability.

\subsection{Dynamical Mass Generation and Semiclassical Quantization}

\begin{figure}[H]
    \centering
    \includegraphics[width=0.7\linewidth]{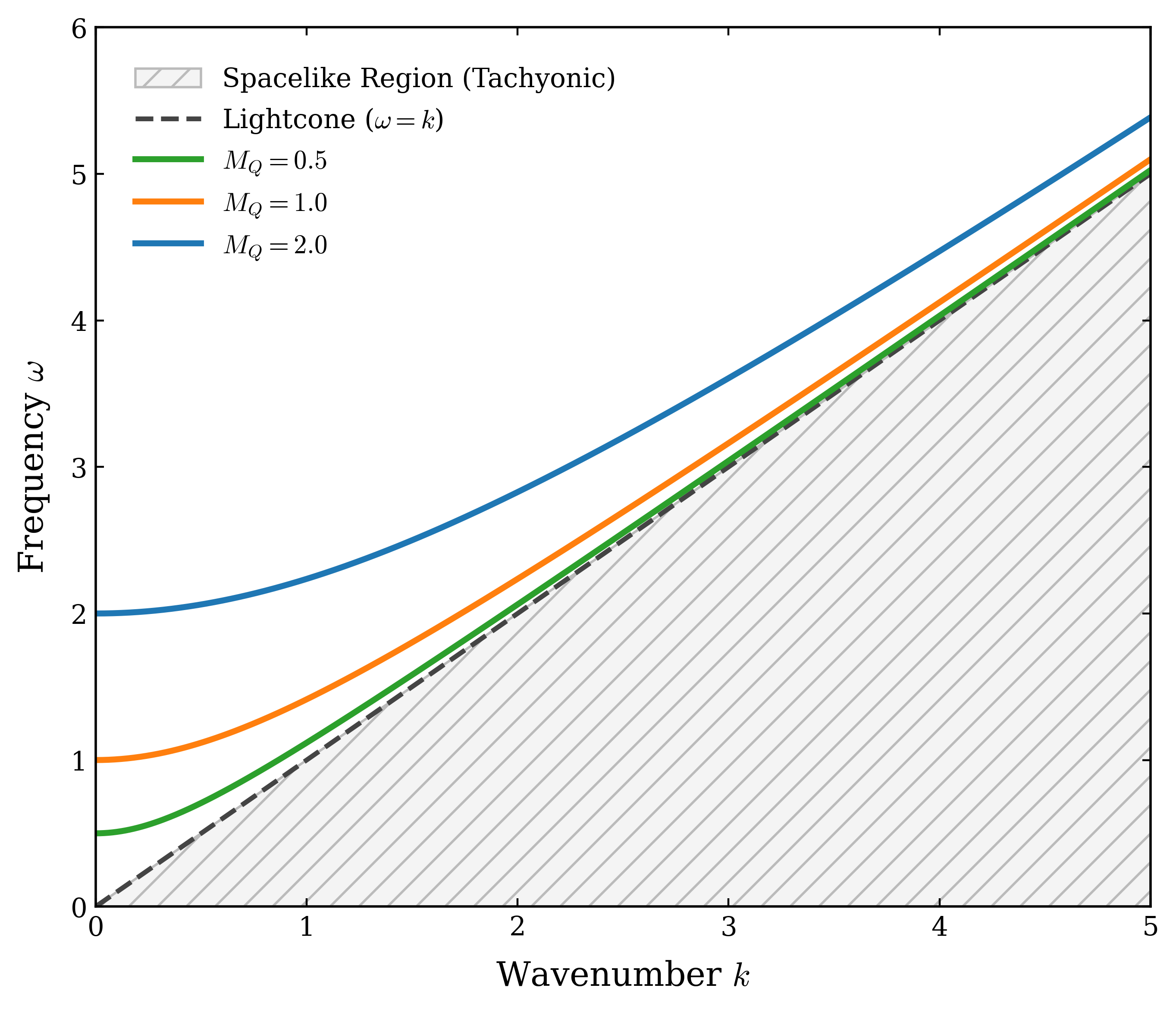}
    \caption{Dispersion phase diagram of the $SU(2)$ coherent states. The dynamically generated mass gap modifies the standard lightcone (black dashed), producing massive relativistic dispersion curves that confine the stable propagating solutions to the timelike region. The spacelike region (grey hatched), which conventionally harbors tachyonic instabilities associated with Savvidy-type vacuum collapse, is excluded by the dynamical infrared cutoff $\omega = k$.}
    \label{fig:phase_space}
\end{figure}

The algebraic decomposition restricts the massive coherent states to the timelike region ($\omega > k$), as depicted in Fig.~\ref{fig:phase_space}. 

The non-linear interaction directly generates an effective classical mass parameter $m_{\text{eff}} \sim (\omega^2 - k^2)^{-1/2}$. In the spacelike region ($\omega < k$), $m_{\text{eff}}^2$ becomes negative, indicating that such modes correspond to tachyonic instabilities rather than stable propagating waves. Thus, the lightcone boundary $\omega = k$ acts as an infrared cutoff.

While $m_{\text{eff}}$ characterizes the classical dispersion, its connection to the quantum mass gap $M_Q$ requires semiclassical quantization. Using the Background Field Method~\cite{Binosi2009}, the partition function is expanded around the classical solution $\bar{A}_\mu$. At the 1-loop level, the quadratic action for the quantum fluctuations $a_\mu$ in the Feynman gauge ($\xi = 1$) is:
\begin{align}
S^{(2)} &= \frac{1}{2} \int d^4x \, a_\mu^a \Big[ -\eta^{\mu\nu} \partial^2 \delta^{ab} - 2 g f^{abc} \bar{A}^{\rho, c} \partial_\rho \eta^{\mu\nu} \nonumber \\
&\quad - g^2 f^{eac} f^{ebd} \bar{A}_\rho^c \bar{A}^{\rho, d} \eta^{\mu\nu} - 2 g f^{abc} \bar{F}^{\mu\nu, c} \Big] a_\nu^b.
\end{align}
The four-point interaction of Yang-Mills theory, evaluated in the macroscopic background $\bar{A}_\mu$, generates a local mass matrix for the quantum gluons: $(M^2)^{ab} = - g^2 f^{eac} f^{ebd} \bar{A}_\rho^c \bar{A}^{\rho, d}$. For the relativistic $SU(2)$ color waves (Section \ref{sec:color_waves}), the background field spans the transverse directions: $\bar{A}_x^1 = f(v)$ and $\bar{A}_y^2 = \kappa f(v)$. Substituting this into the $SU(2)$ structure constants ($f^{abc} = \epsilon^{abc}$), the mass matrix is explicitly diagonalized:
\begin{equation}
(M^2)^{ab} = g^2 f(v)^2 \begin{pmatrix} \kappa^2 & 0 & 0 \\ 0 & 1 & 0 \\ 0 & 0 & 1+\kappa^2 \end{pmatrix}.
\end{equation}
For the dynamically coupled Branch B ($\kappa = 1$), the background imparts a positive mass to all three color components. Since the background is a self-sustaining coherent wave, the physical mass is determined by its spacetime vacuum expectation value $\langle \bar{A}^2 \rangle \equiv \Lambda^2$. In the classical solution, the macroscopic wave amplitude is governed by $f^4 \propto (\omega^2 - k^2)^{-1}$, scaling directly with $m_{\text{eff}}$. Therefore, the pole mass of the quantum gluon (the quantum mass gap $M_Q$) is dynamically generated by the classical condensate:
\begin{equation}
M_Q = g \sqrt{\langle \bar{A}^2 \rangle} \propto g \cdot m_{\text{eff}}.
\end{equation}
This provides an analytical connection between the effective mass parameter of the classical tensor ring and the dynamically generated mass gap in the single-loop quantum propagator, corroborating modern functional renormalization group and Dyson-Schwinger approaches~\cite{Cyrol2018, Huber2020}.

\subsection{Structural Stability of the Algebraic Solutions}

A known instability in constant magnetic background models (e.g., the Savvidy vacuum~\cite{Savvidy1977}) is the Nielsen-Olesen tachyonic instability. For a generic linear fluctuation $a_\mu$, the background field method exposes the second-order stability operator:
\begin{equation}
\mathcal{M}_{\mu\nu}^{ab} a^\nu_b = \left( -\eta_{\mu\nu} (\bar{D}^2)^{ab} - 2 f^{abc} \bar{F}_{\mu\nu}^c \right) a^\nu_b = \lambda a_{\mu}^a.
\end{equation}
In the Savvidy vacuum, the uniform, infinite-volume constant field strength $\bar{F}$ yields a constant, negative spin-magnetic coupling $-2 f^{abc} \bar{F}_{\mu\nu}^c$, which injects a macroscopic tachyonic mass squared ($\lambda < 0$) into specific gluon polarizations, causing exponential vacuum collapse.

The configurations extracted via the algebraic quotient ring method inherently resist this instability through distinct physical mechanisms:

\textbf{1. Sec.~\ref{sec:color_waves}, Branch A (Decoupled Plane Wave):} Because the longitudinal deformation vanishes ($\Omega = 0$), the spin-coupling $-2g\epsilon^{abc}\bar{F}_{\mu\nu}^c$ is sourced purely by the transverse wave. This coupling oscillates in the phase coordinate $v$ with a vanishing time-average, structurally preventing the accumulation of a constant tachyonic mass squared.

\textbf{2. Sec.~\ref{sec:color_waves}, Branches B--C (Coupled Massive Waves):} 
For the fully coupled relativistic branches, the active longitudinal Maurer-Cartan background generates a spacetime-varying field strength. Specifically for Branch B, the geometric cross-terms generate alternating components, yielding a spin-coupling $2 i g [\bar{F}_{\mu\nu}, a^\nu]$ that dynamically modulates the fluctuation modes. Because the background envelope $f(v)$ is a Jacobi elliptic function with zero mean $\langle f(v) \rangle = 0$, the eigenvalues of the spin-coupling matrix purely oscillate. The fluctuation equation maps directly to a matrix Mathieu-type equation:
\begin{equation}
\ddot{a}_{\mu} - \nabla^2 a_{\mu} + \left( M^2(v) \right)_{\mu}^{\ \nu} a_{\nu} = 0.
\end{equation}
Because the mass-squared matrix $M^2(v)$ is positive-definite on average, the coupled waves resist the Savvidy exponential decay.

\textbf{3. Sec.~\ref{sec:static_flux_tubes}, Branch C (Dynamical Meissner Flux Tube):} 
The Savvidy instability inherently afflicts macroscopic, constant magnetic fields. By contrast, the dynamical topological templates derived in Section \ref{sec:static_flux_tubes} confine the magnetic flux. The Artinian asymptotic evaluation guarantees that the background field strengths (e.g., $\bar{F}_{r\phi}, \bar{F}_{rz}$) decay exponentially as modified Bessel functions $K_1(Mr)$. Consequently, the tachyonic spin-magnetic coupling $-2g f^{abc} \bar{F}_{\mu\nu}^c$ is localized to the flux tube core and vanishes in the asymptotic vacuum. Furthermore, the temporal dominance condition dynamically generates a positive-definite asymptotic mass gap squared $M^2 = \Omega_\infty^2 - \tilde{W}_\infty^2 > 0$. This dominant positive mass suppresses any residual long-wavelength tachyonic modes, providing a topological stabilization mechanism.

\textbf{4. Sec.~\ref{sec:su3_dynamics}, Branch B ($SU(3)$ Kinetic Cancellation):}
The stability of the generalized $x^2 y^2$ chaotic oscillator branch is verified via the evaluation of its background fields. The background possesses active spatial components in both the transverse ($R_V$) and longitudinal ($\tilde{K}_V$) directions. Due to the negative spatial metric signature ($\eta_{ii} = -1$), the quantum mass matrix $(M^2)^{ab} = -g^2 f^{eac} f^{ebd} \bar{A}_\rho^c \bar{A}^{\rho, d}$ evaluates to a positive-definite, periodically time-varying mass $M^2(t) \propto g^2 (R_V(t)^2 + \tilde{K}_V(t)^2)$.

Crucially, the Gauss law enforces the kinetic cancellation condition $\tilde{c}_V = -\dot{\theta}_V$. This algebraic locking strips the field strength of its rotational phase derivatives, reducing the commutators to pure amplitude oscillations (e.g., $\bar{F}_{tx} = \dot{R}_V \cos\theta_V T_4 + \dot{R}_V \sin\theta_V T_5$). This has specific spectral consequences:
\begin{enumerate}
    \item The spin-magnetic coupling is purely alternating. Since the amplitude velocity $\dot{R}_V$ is bounded and possesses a vanishing time-average ($\langle \dot{R}_V \rangle = 0$), the tachyonic spin-coupling cannot macroscopically accumulate.
    \item The fluctuation equation for the quantum gluons decouples into a system governed by a positive-definite periodic potential:
    \begin{equation}
    \ddot{a}_i + \left( \mathbf{k}^2 + g^2[R_V(t)^2 + \tilde{K}_V(t)^2] \right) a_i = 0.
    \end{equation}
\end{enumerate}
This maps the quantum fluctuation dynamics onto a stable Hill's equation. The potentially dangerous spin-coupling is dynamically transmuted into a positive-definite oscillating mass, demonstrating that the algebraically extracted $SU(3)$ coherent resonances represent structurally stable classical minima.

\section{Conclusion}
\label{sec:conclusion}

In this paper, an algebraic tensor ring decomposition framework was formulated to extract coherent states from the non-linear equations of Yang-Mills theory. By mapping the differential dynamics into differential-algebraic systems, and evaluating them over specific quotient rings (such as elliptic and Artinian rings), the non-linear PDEs were analyzed. Treating pure-gauge backgrounds as dynamical variables provided the geometric templates required to generate cross-terms and sustain non-linear self-interactions. 

Applying this method, three distinct physical configurations were analyzed:
\begin{enumerate}
    \item \textbf{Relativistic Color Waves (Sec.~\ref{sec:color_waves}):} Evaluated over an elliptic quotient ring, the algebraic ideal bifurcates into three branches. Branch A (Decoupled) yields a standard transverse non-Abelian plane wave with $m_{\text{eff}}=0$. Branch B (Isotropic Coupling) and Branch C (Linear Polarization) both require a non-trivial longitudinal deformation $\Omega \neq 0$ and are confined to the timelike region, demonstrating the dynamical generation of a mass gap as an analytic counterpart to lattice observations~\cite{Athenodorou2021}.
    \item \textbf{Dynamical Flux Tubes (Sec.~\ref{sec:static_flux_tubes}):} A time-dependent helical topological template $U = \exp(n\phi T_3 + kz T_3 - \omega t T_3)$ was introduced to evade Derrick's theorem. The Gauss law ideal performs a topological bifurcation: Branch~A (Coulomb phase) yields unconfined logarithmic potentials; Branch~B (Dyonic phase) establishes a dynamical locking between the azimuthal and longitudinal templates; Branch~C (Symmetric Meissner) admits an Artinian asymptotic analysis yielding Bessel-type exponential screening, stabilized by a temporal dominance condition $M^2 > 0$. This provides a classical realization of the dyonic dual superconductor hypothesis~\cite{Nambu1974, Mandelstam1976, tHooft1981}.
    \item \textbf{SU(3) $x^2y^2$ Resonances (Sec.~\ref{sec:su3_dynamics}):} Activating a spatial Cartan template, the Gauss law ideal bifurcates into four phases. In the non-trivial branches, the kinetic cancellation mechanism enforces $\tilde{c}_V = -\dot{\theta}_V$, linearizing the field strength to purely radial oscillations. The spatial template then promotes the background coupling from a rigid constant to a dynamical field, mapping the amplitude equations onto the $x^2y^2$ chaotic oscillator, with analytical solutions given by Jacobi elliptic functions.
\end{enumerate}

Furthermore, semiclassical analysis (Sec.~\ref{sec:discussion}) indicates that these configurations structurally resist the Savvidy tachyonic instability, either through vanishing time-averaged spin-couplings or by transmuting them into positive-definite oscillating masses. Systematically mapping these classical solutions provides a framework for analyzing non-perturbative dynamics. These configurations can supply the background fields required for semiclassical path-integral quantization and serve as analytical points of comparison for lattice QCD findings. 

The algebraic tensor ring decomposition framework can be generalized. Because the geometric non-linearities in Yang-Mills theory share mathematical similarities with the curvature terms in General Relativity, this algebraic quotient ring technique can be applied to other systems. By mapping local spacetime frames into parameter rings, this methodology offers a potential pathway to investigate the PDEs governing Einstein-Yang-Mills systems, higher-dimensional supergravity, and similar differential geometries.

\end{document}


\title{Supplemental Material for:\\ Systematic Extraction of Exact Yang-Mills Solutions via Algebraic Tensor Ring Decomposition}

\author{Yu-Xuan Zhang}
\affiliation{School of Physics, Nankai University, Tianjin 300071, People's Republic of China}

\author{Jing-Ling Chen}
\email{chenjl@nankai.edu.cn}
\affiliation{Theoretical Physics Division, Chern Institute of Mathematics, Nankai University, Tianjin 300071, People's Republic of China}

\date{\today}

\maketitle

\tableofcontents

\vspace{1cm}

This Supplemental Material provides the exhaustive, step-by-step mathematical derivations supporting the claims made in the main text. It details the explicit calculation of the non-Abelian field strength tensors in locally rotated frames, the algebraic projection of the Yang-Mills equations into differential ideals, the exact quotient ring evaluations, and the semiclassical stability analysis.

\section{Derivations for Relativistic $SU(2)$ Color Waves}
\label{sec:sm_su2_waves}

In this section, we derive the exact equations of motion and the energy-momentum sum rule for the $SU(2)$ color waves presented in Section III of the main text. We adopt the Minkowski metric $\eta_{\mu\nu} = \text{diag}(1, -1, -1, -1)$ and the anti-hermitian generators $T_a = -i\sigma_a/2$, which satisfy the commutation relations $[T_a, T_b] = \epsilon_{abc} T_c$.

The generalized ansatz is defined over the traveling phase coordinate $v = kz - \omega t$. The spacetime derivatives act as $\p_t = -\omega \p_v$ and $\p_z = k \p_v$. The background gauge transformation is $U = \exp(v T_3)$. The locally rotated transverse bases are defined as $T_{1,2}(v) = U^{-1} T_{1,2} U$. Their spacetime derivatives are governed strictly by the commutators with the Maurer-Cartan vacuum connection $A_\mu^{\text{vac}} = U^{-1}\p_\mu U$:
\begin{align}
    A_t^{\text{vac}} &= -\omega T_3, \quad \p_t T_1(v) =[T_1(v), -\omega T_3] = \omega T_2(v), \quad \p_t T_2(v) = [T_2(v), -\omega T_3] = -\omega T_1(v), \label{eq:dt_T} \\
    A_z^{\text{vac}} &= k T_3, \quad \p_z T_1(v) =[T_1(v), k T_3] = -k T_2(v), \quad \p_z T_2(v) = [T_2(v), k T_3] = k T_1(v). \label{eq:dz_T}
\end{align}
The gauge field ansatz is parameterized as:
\begin{equation}
    A_t = c_0(v) T_3, \quad A_z = c_3(v) T_3, \quad A_x = c_1(v) T_1(v), \quad A_y = c_2(v) T_2(v).
\end{equation}

The non-Abelian field strength tensor is $F_{\mu\nu} = \p_\mu A_\nu - \p_\nu A_\mu +[A_\mu, A_\nu]$. We calculate each non-vanishing component explicitly. Prime ($'$) denotes the derivative with respect to $v$.

\textbf{1. The $F_{tx}$ component (Color-electric field in $x$):}
\begin{align}
    \p_t A_x &= \p_t (c_1 T_1) = -\omega c_1' T_1 + c_1 (\omega T_2) = -\omega c_1' T_1 + \omega c_1 T_2, \\
    \p_x A_t &= 0, \\
    [A_t, A_x] &=[c_0 T_3, c_1 T_1] = c_0 c_1[T_3, T_1] = c_0 c_1 T_2.
\end{align}
Summing these terms yields $F_{tx} = -\omega c_1' T_1 + (\omega + c_0) c_1 T_2$. Defining the longitudinal deformation $\Omega(v) = c_0(v) + \omega$, we obtain:
\begin{equation}
    F_{0x} = -F_{tx} = \omega c_1' T_1 - \Omega c_1 T_2.
\end{equation}

\textbf{2. The $F_{ty}$ component (Color-electric field in $y$):}
\begin{align}
    \p_t A_y &= \p_t (c_2 T_2) = -\omega c_2' T_2 + c_2 (-\omega T_1) = -\omega c_2' T_2 - \omega c_2 T_1, \\
    [A_t, A_y] &=[c_0 T_3, c_2 T_2] = c_0 c_2 [T_3, T_2] = -c_0 c_2 T_1.
\end{align}
Summing these yields $F_{ty} = -\omega c_2' T_2 - (\omega + c_0) c_2 T_1$. Thus:
\begin{equation}
    F_{0y} = -F_{ty} = \omega c_2' T_2 + \Omega c_2 T_1.
\end{equation}

\textbf{3. The $F_{tz}$ component (Longitudinal color-electric field):}
\begin{equation}
    F_{tz} = \p_t A_z - \p_z A_t + [A_t, A_z] = -\omega c_3' T_3 - k c_0' T_3 + 0 = -(\omega c_3' + k c_0') T_3.
\end{equation}
Defining $K(v) = c_3(v) - k$, we note that $K' = c_3'$ and $\Omega' = c_0'$. Thus:
\begin{equation}
    F_{0z} = -F_{tz} = (\omega K' + k \Omega') T_3.
\end{equation}

\textbf{4. The $F_{zx}$ and $F_{yz}$ components (Color-magnetic fields):}
\begin{align}
    \p_z A_x &= \p_z (c_1 T_1) = k c_1' T_1 + c_1 (-k T_2) = k c_1' T_1 - k c_1 T_2, \\
    [A_z, A_x] &= [c_3 T_3, c_1 T_1] = c_3 c_1 T_2.
\end{align}
Thus, $F_{zx} = k c_1' T_1 + (c_3 - k) c_1 T_2 = k c_1' T_1 + K c_1 T_2$.
Similarly, for $F_{zy}$:
\begin{align}
    \p_z A_y &= \p_z (c_2 T_2) = k c_2' T_2 + c_2 (k T_1) = k c_2' T_2 + k c_2 T_1, \\
    [A_z, A_y] &= [c_3 T_3, c_2 T_2] = -c_3 c_2 T_1.
\end{align}
Thus, $F_{zy} = k c_2' T_2 - (c_3 - k) c_2 T_1 = k c_2' T_2 - K c_2 T_1$. This gives $F_{yz} = -F_{zy} = -k c_2' T_2 + K c_2 T_1$.

\textbf{5. The $F_{xy}$ component:}
\begin{equation}
    F_{xy} = \p_x A_y - \p_y A_x + [A_x, A_y] = 0 - 0 + [c_1 T_1, c_2 T_2] = c_1 c_2 T_3.
\end{equation}

The Yang-Mills equations are $D_\mu F^{\mu\nu} = \p_\mu F^{\mu\nu} + [A_\mu, F^{\mu\nu}] = 0$. We evaluate the longitudinal components ($\nu = t, z$) to extract the differential ideals. 

For $\nu = t$, the equation is $D_x F^{xt} + D_y F^{yt} + D_z F^{zt} = 0$. Using $F^{xt} = -F_{xt} = F_{0x}$ and $F^{zt} = -F_{zt} = F_{0z}$:
\begin{align}
    D_x F^{xt} &= \p_x F_{0x} + [A_x, F_{0x}] = 0 +[c_1 T_1, \omega c_1' T_1 - \Omega c_1 T_2] = -\Omega c_1^2 [T_1, T_2] = -\Omega c_1^2 T_3, \\
    D_y F^{yt} &= \p_y F_{0y} +[A_y, F_{0y}] = 0 +[c_2 T_2, \omega c_2' T_2 + \Omega c_2 T_1] = \Omega c_2^2[T_2, T_1] = -\Omega c_2^2 T_3, \\
    D_z F^{zt} &= \p_z F_{0z} +[A_z, F_{0z}] = k(\omega K'' + k \Omega'') T_3 +[c_3 T_3, F_{0z} T_3] = k(\omega K'' + k \Omega'') T_3.
\end{align}
Summing these contributions yields the temporal ideal $\mathcal{I}_t$:
\begin{equation}
    \mathcal{I}_t: \quad k(\omega K'' + k \Omega'') - \Omega(c_1^2 + c_2^2) = 0. \label{eq:sm_It}
\end{equation}
For $\nu = z$, the equation is $D_t F^{tz} + D_x F^{xz} + D_y F^{yz} = 0$. Using $F^{tz} = -F_{0z}$, $F^{xz} = F_{zx}$, and $F^{yz} = -F_{zy}$:
\begin{align}
    D_t F^{tz} &= \p_t (-F_{0z}) +[A_t, -F_{0z}] = \omega(\omega K'' + k \Omega'') T_3, \\
    D_x F^{xz} &= [A_x, F_{zx}] =[c_1 T_1, k c_1' T_1 + K c_1 T_2] = K c_1^2 T_3, \\
    D_y F^{yz} &=[A_y, -F_{zy}] =[c_2 T_2, -k c_2' T_2 + K c_2 T_1] = K c_2^2 T_3.
\end{align}
Summing these yields the spatial longitudinal ideal $\mathcal{I}_z$:
\begin{equation}
    \mathcal{I}_z: \quad \omega(\omega K'' + k \Omega'') + K(c_1^2 + c_2^2) = 0. \label{eq:sm_Iz}
\end{equation}
To eliminate the second-derivative term $(\omega K'' + k \Omega'')$, we compute $\omega \times \text{Eq.}(\ref{eq:sm_It}) - k \times \text{Eq.}(\ref{eq:sm_Iz})$:
\begin{equation}
    -(\omega \Omega + k K)(c_1^2 + c_2^2) = 0.
\end{equation}
For any non-trivial transverse wave ($c_1^2 + c_2^2 \neq 0$), this rigorously enforces the kinematic locking relation:
\begin{equation}
    K(v) = -\frac{\omega}{k}\Omega(v). \label{eq:sm_kinematic}
\end{equation}
We now evaluate the transverse Yang-Mills equations. For $\nu = x$, $D_t F^{tx} + D_y F^{yx} + D_z F^{zx} = 0$.
\begin{align}
    D_t F^{tx} &= \p_t (-\omega c_1' T_1 + \Omega c_1 T_2) +[c_0 T_3, -\omega c_1' T_1 + \Omega c_1 T_2] \nonumber \\
               &= \omega^2 c_1'' T_1 - \omega \Omega' c_1 T_2 - \omega \Omega c_1' T_2 - \omega c_0 c_1' T_2 - c_0 \Omega c_1 T_1 \nonumber \\
               &= (\omega^2 c_1'' - c_0 \Omega c_1) T_1 - (\omega \Omega' c_1 + \omega \Omega c_1' + \omega c_0 c_1') T_2.
\end{align}
Using $c_0 = \Omega - \omega$, the $T_1$ component of $D_t F^{tx}$ is $\omega^2 c_1'' - \Omega(\Omega - \omega)c_1$.
\begin{align}
    D_y F^{yx} &= [A_y, -F_{xy}] =[c_2 T_2, -c_1 c_2 T_3] = -c_1 c_2^2 T_1. \\
    D_z F^{zx} &= \p_z (k c_1' T_1 + K c_1 T_2) +[c_3 T_3, k c_1' T_1 + K c_1 T_2] \nonumber \\
               &= k^2 c_1'' T_1 + k K' c_1 T_2 + k K c_1' T_2 + k c_3 c_1' T_2 - c_3 K c_1 T_1.
\end{align}
Using $c_3 = K + k$, the $T_1$ component of $D_z F^{zx}$ is $k^2 c_1'' - K(K + k)c_1$.
Summing the $T_1$ components for the $\nu = x$ equation:
\begin{equation}
    (\omega^2 - k^2) c_1'' - \left[ \Omega(\Omega - \omega) + K(K + k) + c_2^2 \right] c_1 = 0.
\end{equation}
Substituting the kinematic relation $K = -(\omega/k)\Omega$, the term $\Omega(\Omega - \omega) + K(K + k)$ simplifies exactly to $-\frac{\omega^2 - k^2}{k^2} \Omega^2 \equiv -\gamma \Omega^2$. Thus, the $c_1$ equation becomes:
\begin{equation}
    (\omega^2 - k^2) c_1'' + c_1 (c_2^2 + \gamma \Omega^2) = 0. \label{eq:sm_c1_eom}
\end{equation}
By exact symmetry, the $\nu = y$ equation yields:
\begin{equation}
    (\omega^2 - k^2) c_2'' + c_2 (c_1^2 + \gamma \Omega^2) = 0. \label{eq:sm_c2_eom}
\end{equation}
Finally, substituting $K = -(\omega/k)\Omega$ back into the temporal ideal $\mathcal{I}_t$ (Eq.~\ref{eq:sm_It}) yields the $\Omega$ equation:
\begin{equation}
    (\omega^2 - k^2) \Omega'' + \Omega (c_1^2 + c_2^2) = 0. \label{eq:sm_omega_eom}
\end{equation}
We evaluate the system (Eqs.~\ref{eq:sm_c1_eom}-\ref{eq:sm_omega_eom}) over the elliptic quotient ring $\Rellip = \mathbb{R}[f, f'] / \left\langle (f')^2 + \frac{\lambda}{2} f^4 - E_0 \right\rangle$, with the derivation rule $f'' = -\lambda f^3$. We constrain the variables to proportional rays:
\begin{equation}
    c_1(v) = f(v), \quad c_2(v) = \kappa f(v), \quad \Omega(v) = \alpha f(v).
\end{equation}
Substituting these into the equations of motion transforms the differential equations into algebraic polynomials in $f^3$:
\begin{align}
    \text{From } \Omega: \quad & (\omega^2 - k^2) \alpha (-\lambda f^3) + \alpha f (\kappa^2 f^2 + f^2) = 0 \implies \alpha \left[ -\lambda(\omega^2 - k^2) + (1 + \kappa^2) \right] f^3 = 0. \label{eq:sm_ring1} \\
    \text{From } c_1: \quad & (\omega^2 - k^2) (-\lambda f^3) + f (\kappa^2 f^2 + \gamma \alpha^2 f^2) = 0 \implies \left[ -\lambda(\omega^2 - k^2) + (\kappa^2 + \gamma \alpha^2) \right] f^3 = 0. \label{eq:sm_ring2}
\end{align}

\textbf{Branch A (Decoupled Wave):} Assume $\alpha = 0$ (thus $\Omega = 0$). Eq.~(\ref{eq:sm_ring1}) is trivially satisfied. Eq.~(\ref{eq:sm_ring2}) requires $\lambda(\omega^2 - k^2) = \kappa^2$. The $c_2$ equation symmetrically requires $\lambda(\omega^2 - k^2) \kappa = \kappa$. For non-trivial solutions, $\kappa = \pm 1$, yielding $\lambda = 1/(\omega^2 - k^2)$.

\textbf{Branch B (Isotropic Coupled Wave):} Assume $\alpha \neq 0$. Eq.~(\ref{eq:sm_ring1}) demands $\lambda(\omega^2 - k^2) = 1 + \kappa^2$. Eq.~(\ref{eq:sm_ring2}) demands $\lambda(\omega^2 - k^2) = \kappa^2 + \gamma \alpha^2$. Equating the two yields $1 + \kappa^2 = \kappa^2 + \gamma \alpha^2 \implies \gamma \alpha^2 = 1$. Since $\gamma = (\omega^2 - k^2)/k^2$, we find $\alpha^2 = k^2/(\omega^2 - k^2)$. The $c_2$ equation requires $\kappa(1 + \gamma \alpha^2) = \lambda(\omega^2 - k^2)\kappa$. Since $\gamma \alpha^2 = 1$, this gives $2\kappa = (1+\kappa^2)\kappa$, which implies $\kappa = \pm 1$. Consequently, $\lambda = 2/(\omega^2 - k^2)$.

The macroscopic energy density is $\rho = \frac{1}{2}(E^2 + B^2)$. Using the field strengths and $K = -(\omega/k)\Omega$:
\begin{align}
    E^2 &= 2\Tr(F_{0i}F_{0i}^\dagger) = \omega^2((c_1')^2 + (c_2')^2) + \Omega^2(c_1^2 + c_2^2) + (\omega K' + k \Omega')^2 \nonumber \\
        &= \omega^2((c_1')^2 + (c_2')^2) + \Omega^2(c_1^2 + c_2^2) + \frac{(\omega^2 - k^2)^2}{k^2} (\Omega')^2. \\
    B^2 &= 2\Tr(F_{ij}F_{ij}^\dagger) = k^2((c_1')^2 + (c_2')^2) + K^2(c_1^2 + c_2^2) + c_1^2 c_2^2 \nonumber \\
        &= k^2((c_1')^2 + (c_2')^2) + \frac{\omega^2}{k^2}\Omega^2(c_1^2 + c_2^2) + c_1^2 c_2^2.
\end{align}
Summing these and substituting the proportional rays ($c_1=f, c_2=\kappa f, \Omega=\alpha f$) and the ring rule $(f')^2 = E_0 - \frac{\lambda}{2}f^4$:
\begin{equation}
    \rho = \frac{1}{2} \left[ (\omega^2 + k^2)(1+\kappa^2)(f')^2 + \alpha^2 \frac{(\omega^2 - k^2)^2}{k^2} (f')^2 + \alpha^2 \frac{\omega^2 + k^2}{k^2} (1+\kappa^2) f^2 + \kappa^2 f^4 \right].
\end{equation}

\textbf{For Branch A:} $\alpha = 0, \kappa = 1, \lambda = 1/(\omega^2 - k^2)$.
\begin{equation}
    \rho_A = \frac{1}{2} \left[ 2(\omega^2 + k^2)(f')^2 + f^4 \right] = (\omega^2 + k^2) \left( E_0 - \frac{1}{2(\omega^2 - k^2)}f^4 \right) + \frac{1}{2}f^4 = (\omega^2 + k^2)E_0 - \frac{k^2}{\omega^2 - k^2}f^4.
\end{equation}

\textbf{For Branch B:} $\alpha^2 = k^2/(\omega^2 - k^2), \kappa = 1, \lambda = 2/(\omega^2 - k^2)$.
\begin{align}
    \rho_B &= \frac{1}{2} \left[ 2(\omega^2 + k^2)(f')^2 + (\omega^2 - k^2)(f')^2 + 2\frac{\omega^2 + k^2}{\omega^2 - k^2} f^2 + f^4 \right] \nonumber \\
           &= \frac{1}{2} \left[ (3\omega^2 + k^2)(f')^2 + \frac{3\omega^2 + k^2}{\omega^2 - k^2} f^4 \right] \nonumber \\
           &= \frac{1}{2}(3\omega^2 + k^2) \left( E_0 - \frac{1}{\omega^2 - k^2}f^4 \right) + \frac{1}{2}\frac{3\omega^2 + k^2}{\omega^2 - k^2} f^4 = \frac{1}{2}(3\omega^2 + k^2)E_0.
\end{align}
The $f^4$ terms cancel exactly, proving that the macroscopic energy density of the isotropically coupled wave is strictly constant in spacetime.

\section{Derivations for Dynamical Helical Flux Tubes}
\label{sec:sm_flux_tubes}

In this section, we provide the detailed derivation of the differential ideals and the Artinian asymptotic truncation for the cylindrical flux tubes presented in Section IV. The metric is $ds^2 = dt^2 - dr^2 - r^2 d\phi^2 - dz^2$. The non-zero contravariant metric components are $g^{tt}=1, g^{rr}=-1, g^{\phi\phi}=-1/r^2, g^{zz}=-1$. The metric determinant is $\sqrt{-g} = r$.

The background transformation is $U = \exp \left[ (n\phi + kz - \omega t)T_3 \right]$. The rotated bases $T_{1,2}(t, \phi, z) = U^{-1} T_{1,2} U$ satisfy:
\begin{equation}
    \p_t T_1 = \omega T_2, \quad \p_\phi T_1 = -n T_2, \quad \p_z T_1 = -k T_2.
\end{equation}
The generalized ansatz is:
\begin{equation}
    A_t = K(r) T_3, \quad A_\phi = \Phi(r) T_3 + r c_1(r) T_1, \quad A_z = W(r) T_3 + c_2(r) T_1.
\end{equation}
We define the deformations $\Omega(r) = K(r) + \omega$, $\tilde{\Phi}(r) = \Phi(r) - n$, and $\tilde{W}(r) = W(r) - k$.

We calculate the covariant field strength tensor $F_{\mu\nu} = \p_\mu A_\nu - \p_\nu A_\mu + [A_\mu, A_\nu]$.

\textbf{1. The $F_{tr}$ component:}
\begin{equation}
    F_{tr} = \p_t A_r - \p_r A_t +[A_t, A_r] = 0 - K' T_3 + 0 = -K' T_3 = -\Omega' T_3.
\end{equation}

\textbf{2. The $F_{t\phi}$ component:}
\begin{align}
    \p_t A_\phi &= \p_t (r c_1 T_1) = \omega r c_1 T_2, \\
    \p_\phi A_t &= 0, \\
    [A_t, A_\phi] &=[K T_3, \Phi T_3 + r c_1 T_1] = K r c_1[T_3, T_1] = K r c_1 T_2.
\end{align}
Summing these yields $F_{t\phi} = (\omega + K) r c_1 T_2 = \Omega r c_1 T_2$.

\textbf{3. The $F_{tz}$ component:}
\begin{align}
    \p_t A_z &= \p_t (c_2 T_1) = \omega c_2 T_2, \\
    [A_t, A_z] &= [K T_3, W T_3 + c_2 T_1] = K c_2 T_2.
\end{align}
Summing these yields $F_{tz} = (\omega + K) c_2 T_2 = \Omega c_2 T_2$.

\textbf{4. The $F_{r\phi}$ component:}
\begin{equation}
    F_{r\phi} = \p_r A_\phi - \p_\phi A_r +[A_r, A_\phi] = \Phi' T_3 + (r c_1)' T_1 = \tilde{\Phi}' T_3 + (r c_1)' T_1.
\end{equation}

\textbf{5. The $F_{rz}$ component:}
\begin{equation}
    F_{rz} = \p_r A_z - \p_z A_r +[A_r, A_z] = W' T_3 + c_2' T_1 = \tilde{W}' T_3 + c_2' T_1.
\end{equation}

\textbf{6. The $F_{\phi z}$ component:}
\begin{align}
    \p_\phi A_z &= \p_\phi (c_2 T_1) = -n c_2 T_2, \\
    \p_z A_\phi &= \p_z (r c_1 T_1) = -k r c_1 T_2, \\
    [A_\phi, A_z] &=[\Phi T_3 + r c_1 T_1, W T_3 + c_2 T_1] = \Phi c_2 T_2 - r c_1 W T_2.
\end{align}
Summing these yields $F_{\phi z} = (\Phi - n) c_2 T_2 - r c_1 (W - k) T_2 = (\tilde{\Phi} c_2 - r c_1 \tilde{W}) T_2$.

The Gauss law is $D_\mu F^{\mu t} = \frac{1}{\sqrt{-g}} \p_\mu (\sqrt{-g} F^{\mu t}) + [A_\mu, F^{\mu t}] = 0$. We evaluate each term.

\textbf{Radial term ($D_r F^{rt}$):}
\begin{equation}
    F^{rt} = g^{rr}g^{tt} F_{rt} = (-1)(1) (-\Omega' T_3) = \Omega' T_3.
\end{equation}
\begin{equation}
    D_r F^{rt} = \frac{1}{r} \p_r (r \Omega' T_3) = \left( \Omega'' + \frac{1}{r}\Omega' \right) T_3.
\end{equation}

\textbf{Azimuthal term ($D_\phi F^{\phi t}$):}
\begin{equation}
    F^{\phi t} = g^{\phi\phi}g^{tt} F_{\phi t} = \left(-\frac{1}{r^2}\right)(1) (-\Omega r c_1 T_2) = \frac{\Omega c_1}{r} T_2.
\end{equation}
\begin{align}
    \p_\phi F^{\phi t} &= \frac{\Omega c_1}{r} \p_\phi T_2 = \frac{\Omega c_1}{r} (n T_1), \\
    [A_\phi, F^{\phi t}] &= \left[ \Phi T_3 + r c_1 T_1, \frac{\Omega c_1}{r} T_2 \right] = \Phi \frac{\Omega c_1}{r} (-T_1) + r c_1 \frac{\Omega c_1}{r} (T_3) = -\frac{\Phi \Omega c_1}{r} T_1 + \Omega c_1^2 T_3.
\end{align}
Summing these yields $D_\phi F^{\phi t} = -\frac{\Omega c_1}{r}(\Phi - n) T_1 + \Omega c_1^2 T_3 = -\frac{\Omega c_1}{r}\tilde{\Phi} T_1 + \Omega c_1^2 T_3$.

\textbf{Longitudinal term ($D_z F^{zt}$):}
\begin{equation}
    F^{zt} = g^{zz}g^{tt} F_{zt} = (-1)(1) (-\Omega c_2 T_2) = \Omega c_2 T_2.
\end{equation}
\begin{align}
    \p_z F^{zt} &= \Omega c_2 \p_z T_2 = \Omega c_2 (k T_1), \\
    [A_z, F^{zt}] &= [W T_3 + c_2 T_1, \Omega c_2 T_2] = -W \Omega c_2 T_1 + \Omega c_2^2 T_3.
\end{align}
Summing these yields $D_z F^{zt} = -\Omega c_2 (W - k) T_1 + \Omega c_2^2 T_3 = -\Omega c_2 \tilde{W} T_1 + \Omega c_2^2 T_3$.

Summing all three terms, the $T_3$ component gives the differential equation for $\Omega$, while the $T_1$ component must vanish identically, yielding the primary Gauss ideal:
\begin{equation}
    \Igauss: \quad -\frac{\Omega}{r} \left( \tilde{\Phi} c_1 + r \tilde{W} c_2 \right) T_1 = 0.
\end{equation}
The radial Ampère law is $D_\mu F^{\mu r} = D_t F^{tr} + D_\phi F^{\phi r} + D_z F^{zr} = 0$. We isolate the $T_2$ cross-terms to find the secondary ideal.

\textbf{Temporal term ($D_t F^{tr}$):}
\begin{equation}
    F^{tr} = -F^{rt} = -\Omega' T_3. \quad D_t F^{tr} = \p_t (-\Omega' T_3) + [K T_3, -\Omega' T_3] = 0.
\end{equation}

\textbf{Azimuthal term ($D_\phi F^{\phi r}$):}
\begin{equation}
    F^{\phi r} = g^{\phi\phi}g^{rr} F_{\phi r} = \left(-\frac{1}{r^2}\right)(-1) (-F_{r\phi}) = -\frac{1}{r^2} \left( \tilde{\Phi}' T_3 + (r c_1)' T_1 \right).
\end{equation}
The $T_2$ contribution from $\p_\phi F^{\phi r}$ is $-\frac{1}{r^2} (r c_1)' \p_\phi T_1 = \frac{n}{r^2} (r c_1)' T_2$. \\
The $T_2$ contribution from $[A_\phi, F^{\phi r}]$ is:
\begin{equation}
    \left[ \Phi T_3, -\frac{1}{r^2} (r c_1)' T_1 \right] + \left[ r c_1 T_1, -\frac{1}{r^2} \tilde{\Phi}' T_3 \right] = -\frac{\Phi}{r^2} (r c_1)' T_2 + \frac{c_1}{r} \tilde{\Phi}' T_2.
\end{equation}
Total $T_2$ from $D_\phi F^{\phi r}$:
\begin{equation}
    \frac{c_1}{r} \tilde{\Phi}' - \frac{\Phi - n}{r^2} (r c_1)' = \frac{c_1}{r} \tilde{\Phi}' - \frac{\tilde{\Phi}}{r^2} (r c_1)' = c_1^2 \p_r \left( \frac{\tilde{\Phi}}{r c_1} \right) T_2.
\end{equation}

\textbf{Longitudinal term ($D_z F^{zr}$):}
\begin{equation}
    F^{zr} = g^{zz}g^{rr} F_{zr} = (-1)(-1) (-F_{rz}) = -(\tilde{W}' T_3 + c_2' T_1).
\end{equation}
The $T_2$ contribution from $\p_z F^{zr}$ is:
\begin{equation}
    -c_2' \p_z T_1 = k c_2' T_2. 
\end{equation}
The $T_2$ contribution from $[A_z, F^{zr}]$ is:
\begin{equation}[W T_3, -c_2' T_1] +[c_2 T_1, -\tilde{W}' T_3] = -W c_2' T_2 + c_2 \tilde{W}' T_2.
\end{equation}
Total $T_2$ from $D_z F^{zr}$:
\begin{equation}
    c_2 \tilde{W}' - (W - k) c_2' = c_2 \tilde{W}' - \tilde{W} c_2' = c_2^2 \p_r \left( \frac{\tilde{W}}{c_2} \right) T_2.
\end{equation}

Summing these and multiplying by $r^2$ yields the exact secondary ideal $\Iradial$:
\begin{equation}
    \Iradial: \quad (r c_1)^2 \left( \frac{\tilde{\Phi}}{r c_1} \right)' + r^2 c_2^2 \left( \frac{\tilde{W}}{c_2} \right)' = 0.
\end{equation}
For Branch C, we set $c_2 = 0$. The primary ideal $\Igauss$ enforces $\tilde{\Phi} = 0$. The $T_3$ component of the Gauss law (derived above) becomes:
\begin{equation}
    \Omega'' + \frac{1}{r}\Omega' + c_1^2 \Omega = 0.
\end{equation}
The $T_3$ component of the longitudinal Ampère law ($D_\mu F^{\mu z} = 0$) symmetrically yields:
\begin{equation}
    \tilde{W}'' + \frac{1}{r}\tilde{W}' - c_1^2 \tilde{W} = 0.
\end{equation}
The azimuthal Ampère law ($D_\mu F^{\mu \phi} = 0$) evaluated for the $T_1$ component yields the condensate equation:
\begin{equation}
    c_1'' + \frac{1}{r} c_1' - \frac{1}{r^2} c_1 - (\Omega^2 - \tilde{W}^2) c_1 = 0.
\end{equation}
To extract the asymptotic behavior ($r \to \infty$), we evaluate the system over an Artinian quotient ring defined by the nilpotent ideal $\langle c_1^2 \rangle = 0$. In this ring, the non-linear terms $c_1^2 \Omega$ and $c_1^2 \tilde{W}$ vanish identically. The longitudinal equations strictly truncate to Laplace equations:
\begin{equation}
    \Omega'' + \frac{1}{r}\Omega' = 0, \quad \tilde{W}'' + \frac{1}{r}\tilde{W}' = 0.
\end{equation}
To maintain finite energy, the potentials must asymptote to constants: $\Omega(r) \to \Omega_\infty$ and $\tilde{W}(r) \to \tilde{W}_\infty$. Substituting these constants back into the condensate equation yields a linear ODE:
\begin{equation}
    c_1'' + \frac{1}{r} c_1' - \left( M^2 + \frac{1}{r^2} \right) c_1 = 0, \quad \text{where } M^2 = \Omega_\infty^2 - \tilde{W}_\infty^2.
\end{equation}
This is exactly the modified Bessel differential equation of order 1. Provided the temporal dominance condition $M^2 > 0$ holds, the unique normalizable solution in the Artinian ring is $c_1(r) \propto K_1(Mr)$, rigorously proving the exponential screening mechanism.

\section{Derivations for $SU(3)$ Dynamics and Kinetic Cancellation}
\label{sec:sm_su3_dynamics}

In this section, we provide the rigorous algebraic proof of the kinetic cancellation mechanism that maps the highly coupled $SU(3)$ Yang-Mills equations onto the generalized $x^2y^2$ chaotic oscillator (Section V of the main text). We explicitly demonstrate why the Gauss law yields an algebraic ideal, whereas the spatial Ampère laws yield the dynamical ordinary differential equations (ODEs).

We utilize the standard anti-hermitian generators for $\mathfrak{su}(3)$. The Cartan subalgebra is spanned by $T_3$ and $T_8$. The V-spin sector is spanned by $T_4, T_5$, and the U-spin sector by $T_6, T_7$. The relevant commutation relations connecting the Cartan generators to the V-spin sector are:
\begin{align}
    [T_3, T_4] &= \frac{1}{2} T_5, \quad [T_3, T_5] = -\frac{1}{2} T_4, \label{eq:sm_comm_T3} \\
    [T_8, T_4] &= \frac{\sqrt{3}}{2} T_5, \quad [T_8, T_5] = -\frac{\sqrt{3}}{2} T_4, \label{eq:sm_comm_T8} \\
    [T_4, T_5] &= \frac{1}{2} T_3 + \frac{\sqrt{3}}{2} T_8 \equiv Y_V. \label{eq:sm_comm_YV}
\end{align}
Symmetric relations hold for the U-spin sector with $Y_U \equiv -\frac{1}{2} T_3 + \frac{\sqrt{3}}{2} T_8$.

The temporal Cartan field is parameterized as $A_t = c_0(t) T_3 + \Gamma(t) T_8$. The V-spin transverse field is $A_x = c_1(t) T_4 + c_2(t) T_5$. We define the effective temporal Cartan coupling as:
\begin{equation}
    \tilde{c}_V(t) = \frac{1}{2} c_0(t) + \frac{\sqrt{3}}{2} \Gamma(t).
\end{equation}
Using Eqs.~(\ref{eq:sm_comm_T3}) and (\ref{eq:sm_comm_T8}), the commutator $[A_t, A_x]$ evaluates exactly to:
\begin{align}
    [A_t, T_4] &= c_0 [T_3, T_4] + \Gamma[T_8, T_4] = \left( \frac{1}{2} c_0 + \frac{\sqrt{3}}{2} \Gamma \right) T_5 = \tilde{c}_V T_5, \\
    [A_t, T_5] &= c_0 [T_3, T_5] + \Gamma [T_8, T_5] = \left( -\frac{1}{2} c_0 - \frac{\sqrt{3}}{2} \Gamma \right) T_4 = -\tilde{c}_V T_4.
\end{align}
Therefore, the commutator between the temporal and spatial fields is:
\begin{equation}
    [A_x, A_t] = -[A_t, A_x] = -\left( c_1 \tilde{c}_V T_5 - c_2 \tilde{c}_V T_4 \right) = c_2 \tilde{c}_V T_4 - c_1 \tilde{c}_V T_5. \label{eq:sm_AxAt}
\end{equation}
Because the field tensor is antisymmetric ($F_{tt} = 0$), the Gauss law ($D_i F^{it} = 0$) lacks second-order time derivatives ($\ddot{A}_t$). Consequently, $A_t$ acts as a Lagrange multiplier, and the Gauss law strictly projects out an algebraic constraint rather than a dynamical evolution equation.

Since the fields depend only on time, the spatial derivatives vanish ($\p_x A_t = 0$). The field strength $F_{xt} = \p_x A_t - \p_t A_x + [A_x, A_t]$ becomes:
\begin{equation}
    F_{xt} = -(\dot{c}_1 T_4 + \dot{c}_2 T_5) + (c_2 \tilde{c}_V T_4 - c_1 \tilde{c}_V T_5) = -(\dot{c}_1 - \tilde{c}_V c_2) T_4 - (\dot{c}_2 + \tilde{c}_V c_1) T_5.
\end{equation}
The Gauss law requires $D_x F^{xt} = \p_x F^{xt} + [A_x, F^{xt}] = 0$. Since $\p_x F^{xt} = 0$, we evaluate the commutator:
\begin{align}[A_x, F^{xt}] &=[c_1 T_4 + c_2 T_5, -(\dot{c}_1 - \tilde{c}_V c_2) T_4 - (\dot{c}_2 + \tilde{c}_V c_1) T_5] \nonumber \\
                  &= -c_1 (\dot{c}_2 + \tilde{c}_V c_1)[T_4, T_5] - c_2 (\dot{c}_1 - \tilde{c}_V c_2) [T_5, T_4].
\end{align}
Using $[T_5, T_4] = -[T_4, T_5] = -Y_V$, this simplifies to:
\begin{equation}
    D_x F^{xt} = \left[ c_2 (\dot{c}_1 - \tilde{c}_V c_2) - c_1 (\dot{c}_2 + \tilde{c}_V c_1) \right] Y_V = \left[ (c_2 \dot{c}_1 - c_1 \dot{c}_2) - \tilde{c}_V (c_1^2 + c_2^2) \right] Y_V = 0.
\end{equation}
We substitute the polar coordinates $c_1 = R_V \cos\theta_V$ and $c_2 = R_V \sin\theta_V$. Their time derivatives are:
\begin{align}
    \dot{c}_1 &= \dot{R}_V \cos\theta_V - R_V \dot{\theta}_V \sin\theta_V, \\
    \dot{c}_2 &= \dot{R}_V \sin\theta_V + R_V \dot{\theta}_V \cos\theta_V.
\end{align}
The angular momentum term evaluates to:
\begin{align}
    c_2 \dot{c}_1 - c_1 \dot{c}_2 &= R_V \sin\theta_V (\dot{R}_V \cos\theta_V - R_V \dot{\theta}_V \sin\theta_V) - R_V \cos\theta_V (\dot{R}_V \sin\theta_V + R_V \dot{\theta}_V \cos\theta_V) \nonumber \\
                                  &= -R_V^2 \dot{\theta}_V (\sin^2\theta_V + \cos^2\theta_V) = -R_V^2 \dot{\theta}_V.
\end{align}
Substituting this back into the Gauss law yields the exact primary ideal:
\begin{equation}
    D_x F^{xt} = \left( -R_V^2 \dot{\theta}_V - \tilde{c}_V R_V^2 \right) Y_V = -R_V^2 (\dot{\theta}_V + \tilde{c}_V) Y_V = 0.
\end{equation}
Including the symmetric U-spin sector, we obtain the full Gauss ideal $\Igauss$ presented in Eq. (46) of the main text. For a non-trivial transverse amplitude ($R_V \neq 0$), this rigorously enforces the kinetic cancellation condition $\tilde{c}_V = -\dot{\theta}_V$.

Unlike the Gauss law, the spatial Ampère laws ($D_\mu F^{\mu i} = 0$) contain second-order time derivatives ($\ddot{A}_i$). Because the $SU(3)$ ansatz is reduced to a 1D mechanical system (spatially uniform), the spatial derivatives vanish, and the Ampère laws directly yield the dynamical ODEs without generating secondary algebraic constraints.

We apply the kinetic cancellation condition $\tilde{c}_V = -\dot{\theta}_V$ to the field strength $F_{tx} = -F_{xt}$:
\begin{equation}
    F_{tx} = (\dot{c}_1 - \tilde{c}_V c_2) T_4 + (\dot{c}_2 + \tilde{c}_V c_1) T_5.
\end{equation}
Substituting the polar derivatives and $\tilde{c}_V = -\dot{\theta}_V$:
\begin{align}
    \dot{c}_1 - \tilde{c}_V c_2 &= (\dot{R}_V \cos\theta_V - R_V \dot{\theta}_V \sin\theta_V) - (-\dot{\theta}_V)(R_V \sin\theta_V) = \dot{R}_V \cos\theta_V, \\
    \dot{c}_2 + \tilde{c}_V c_1 &= (\dot{R}_V \sin\theta_V + R_V \dot{\theta}_V \cos\theta_V) + (-\dot{\theta}_V)(R_V \cos\theta_V) = \dot{R}_V \sin\theta_V.
\end{align}
Thus, the field strength is stripped of all rotational phase derivatives, reducing to pure amplitude oscillations:
\begin{equation}
    F_{tx} = \dot{R}_V \cos\theta_V T_4 + \dot{R}_V \sin\theta_V T_5.
\end{equation}
Next, we evaluate the covariant time derivative $D_t F^{tx} = \p_t F^{tx} + [A_t, F^{tx}]$.
\begin{equation}
    \p_t F^{tx} = (\ddot{R}_V \cos\theta_V - \dot{R}_V \dot{\theta}_V \sin\theta_V) T_4 + (\ddot{R}_V \sin\theta_V + \dot{R}_V \dot{\theta}_V \cos\theta_V) T_5.
\end{equation}
The commutator is:
\begin{align}
    [A_t, F^{tx}] &= \dot{R}_V \cos\theta_V [A_t, T_4] + \dot{R}_V \sin\theta_V[A_t, T_5] \nonumber \\
                  &= \dot{R}_V \cos\theta_V (\tilde{c}_V T_5) + \dot{R}_V \sin\theta_V (-\tilde{c}_V T_4).
\end{align}
Substituting $\tilde{c}_V = -\dot{\theta}_V$:
\begin{equation}[A_t, F^{tx}] = -\dot{R}_V \dot{\theta}_V \cos\theta_V T_5 + \dot{R}_V \dot{\theta}_V \sin\theta_V T_4.
\end{equation}
Adding $\p_t F^{tx}$ and $[A_t, F^{tx}]$, the terms containing $\dot{\theta}_V$ cancel exactly:
\begin{equation}
    D_t F^{tx} = \ddot{R}_V \cos\theta_V T_4 + \ddot{R}_V \sin\theta_V T_5. \label{eq:sm_DtFtx}
\end{equation}
Finally, we evaluate the spatial contribution $D_z F^{zx} = \p_z F^{zx} + [A_z, F^{zx}]$. Since the fields are spatially uniform, $\p_z F^{zx} = 0$. The field strength is $F_{zx} =[A_z, A_x]$. With $A_z = K_1 T_3 + K_2 T_8$, we define the effective spatial Cartan coupling $\tilde{K}_V = \frac{1}{2}K_1 + \frac{\sqrt{3}}{2}K_2$. Analogous to Eq.~(\ref{eq:sm_AxAt}), we have:
\begin{equation}
    F_{zx} =[A_z, A_x] = c_1 \tilde{K}_V T_5 - c_2 \tilde{K}_V T_4.
\end{equation}
The covariant derivative is:
\begin{align}
    D_z F^{zx} &=[A_z, c_1 \tilde{K}_V T_5 - c_2 \tilde{K}_V T_4] = c_1 \tilde{K}_V [A_z, T_5] - c_2 \tilde{K}_V [A_z, T_4] \nonumber \\
               &= c_1 \tilde{K}_V (-\tilde{K}_V T_4) - c_2 \tilde{K}_V (\tilde{K}_V T_5) = -\tilde{K}_V^2 (c_1 T_4 + c_2 T_5) \nonumber \\
               &= -\tilde{K}_V^2 R_V (\cos\theta_V T_4 + \sin\theta_V T_5). \label{eq:sm_DzFzx}
\end{align}
The spatial Ampère law for the $x$-direction is $D_t F^{tx} + D_z F^{zx} = 0$. Combining Eq.~(\ref{eq:sm_DtFtx}) and Eq.~(\ref{eq:sm_DzFzx}):
\begin{equation}
    (\ddot{R}_V + \tilde{K}_V^2 R_V) (\cos\theta_V T_4 + \sin\theta_V T_5) = 0.
\end{equation}
Factoring out the linearly independent generators yields the exact $x^2y^2$ oscillator equation for Branch B:
\begin{equation}
    \ddot{R}_V + \tilde{K}_V^2 R_V = 0.
\end{equation}
For Branch D, both V-spin and U-spin sectors are active. The spatial Ampère law for the $z$-direction is $D_t F^{tz} + D_x F^{xz} + D_y F^{yz} = 0$. 

First, $F_{tz} = \p_t A_z - \p_z A_t + [A_t, A_z] = \dot{K}_1 T_3 + \dot{K}_2 T_8$. Its covariant time derivative is simply $D_t F^{tz} = \ddot{K}_1 T_3 + \ddot{K}_2 T_8$.

Next, we evaluate $D_x F^{xz} =[A_x, -F_{zx}]$. Using $F_{zx} = c_1 \tilde{K}_V T_5 - c_2 \tilde{K}_V T_4$:
\begin{align}[A_x, -F_{zx}] &=[c_1 T_4 + c_2 T_5, -c_1 \tilde{K}_V T_5 + c_2 \tilde{K}_V T_4] \nonumber \\
                   &= -c_1^2 \tilde{K}_V[T_4, T_5] + c_2^2 \tilde{K}_V[T_5, T_4] = -(c_1^2 + c_2^2) \tilde{K}_V Y_V = -R_V^2 \tilde{K}_V \left( \frac{1}{2} T_3 + \frac{\sqrt{3}}{2} T_8 \right).
\end{align}
Symmetrically, for the U-spin sector, $D_y F^{yz} = -R_U^2 \tilde{K}_U Y_U = -R_U^2 \tilde{K}_U \left( -\frac{1}{2} T_3 + \frac{\sqrt{3}}{2} T_8 \right)$.

Summing these contributions for the $z$-direction Ampère law:
\begin{equation}
    (\ddot{K}_1 T_3 + \ddot{K}_2 T_8) - R_V^2 \tilde{K}_V \left( \frac{1}{2} T_3 + \frac{\sqrt{3}}{2} T_8 \right) - R_U^2 \tilde{K}_U \left( -\frac{1}{2} T_3 + \frac{\sqrt{3}}{2} T_8 \right) = 0.
\end{equation}
Separating the linearly independent Cartan generators $T_3$ and $T_8$ yields the exact coupled equations for the spatial Cartan fields (Eqs. 52-53 in the main text):
\begin{align}
    \ddot{K}_1 - \frac{1}{2} R_V^2 \tilde{K}_V + \frac{1}{2} R_U^2 \tilde{K}_U &= 0, \\
    \ddot{K}_2 - \frac{\sqrt{3}}{2} R_V^2 \tilde{K}_V - \frac{\sqrt{3}}{2} R_U^2 \tilde{K}_U &= 0.
\end{align}

\section{Derivations for Semiclassical Quantization}
\label{sec:sm_stability}

In Section VI of the main text, the quantum mass matrix for the gluon fluctuations is defined via the Background Field Method. We provide the explicit diagonalization for $SU(2)$ and the rigorous proof of positive-definiteness for $SU(3)$.

\subsection{SU(2) Background Mass Matrix Diagonalization}

The local mass matrix generated by the four-point interaction in the macroscopic background $\bar{A}_\mu$ is:
\begin{equation}
    (M^2)^{ab} = -g^2 f^{eac} f^{ebd} \bar{A}_\rho^c \bar{A}^{\rho, d}.
\end{equation}
For the $SU(2)$ color waves, the background field spans the spatial transverse directions: $\bar{A}_x^1 = f(v)$ and $\bar{A}_y^2 = \kappa f(v)$. Using the Minkowski metric $\eta_{\mu\nu} = \text{diag}(1, -1, -1, -1)$, the contravariant components are $\bar{A}^x_c = -\bar{A}_x^c$ and $\bar{A}^y_c = -\bar{A}_y^c$. The contraction over the Lorentz index $\rho$ yields:
\begin{equation}
    \bar{A}_\rho^c \bar{A}^{\rho, d} = \bar{A}_x^c \bar{A}^x_d + \bar{A}_y^c \bar{A}^y_d = - (\bar{A}_x^c \bar{A}_x^d + \bar{A}_y^c \bar{A}_y^d).
\end{equation}
Substituting this and the $SU(2)$ structure constants $f^{abc} = \epsilon^{abc}$, the mass matrix becomes strictly positive-definite:
\begin{equation}
    (M^2)^{ab} = g^2 \epsilon^{eac} \epsilon^{ebd} (\bar{A}_x^c \bar{A}_x^d + \bar{A}_y^c \bar{A}_y^d).
\end{equation}
We utilize the standard Levi-Civita identity $\epsilon^{eac} \epsilon^{ebd} = \epsilon^{ace} \epsilon^{bde} = \delta^{ab} \delta^{cd} - \delta^{ad} \delta^{bc}$. Substituting this identity:
\begin{align}
    (M^2)^{ab} &= g^2 (\delta^{ab} \delta^{cd} - \delta^{ad} \delta^{bc}) (\bar{A}_x^c \bar{A}_x^d + \bar{A}_y^c \bar{A}_y^d) \nonumber \\
               &= g^2 \left[ \delta^{ab} (\bar{A}_x^c \bar{A}_x^c + \bar{A}_y^c \bar{A}_y^c) - (\bar{A}_x^b \bar{A}_x^a + \bar{A}_y^b \bar{A}_y^a) \right].
\end{align}
We evaluate the trace term $\bar{A}_x^c \bar{A}_x^c + \bar{A}_y^c \bar{A}_y^c$. Since $\bar{A}_x$ is non-zero only for $c=1$ and $\bar{A}_y$ is non-zero only for $c=2$:
\begin{equation}
    \bar{A}_x^c \bar{A}_x^c + \bar{A}_y^c \bar{A}_y^c = (\bar{A}_x^1)^2 + (\bar{A}_y^2)^2 = f^2 + \kappa^2 f^2 = f^2(1 + \kappa^2).
\end{equation}
Now we evaluate the components of $(M^2)^{ab}$ explicitly. For the diagonal components ($a = b$):
\begin{align}
    (M^2)^{11} &= g^2 \left[ 1 \cdot f^2(1 + \kappa^2) - ((\bar{A}_x^1)^2 + (\bar{A}_y^1)^2) \right] = g^2 \left[ f^2(1 + \kappa^2) - (f^2 + 0) \right] = g^2 \kappa^2 f^2, \\
    (M^2)^{22} &= g^2 \left[ 1 \cdot f^2(1 + \kappa^2) - ((\bar{A}_x^2)^2 + (\bar{A}_y^2)^2) \right] = g^2 \left[ f^2(1 + \kappa^2) - (0 + \kappa^2 f^2) \right] = g^2 f^2, \\
    (M^2)^{33} &= g^2 \left[ 1 \cdot f^2(1 + \kappa^2) - ((\bar{A}_x^3)^2 + (\bar{A}_y^3)^2) \right] = g^2 \left[ f^2(1 + \kappa^2) - (0 + 0) \right] = g^2 (1 + \kappa^2) f^2.
\end{align}
For the off-diagonal components ($a \neq b$), the first term $\delta^{ab}$ vanishes. The second term is $-g^2 (\bar{A}_x^b \bar{A}_x^a + \bar{A}_y^b \bar{A}_y^a)$. Since $\bar{A}_x$ is strictly aligned with color index 1 and $\bar{A}_y$ with color index 2, there is no pair $(a, b)$ with $a \neq b$ where both fields are simultaneously non-zero. Thus, all cross-terms vanish identically.

The mass matrix is exactly diagonalized:
\begin{equation}
    (M^2)^{ab} = g^2 f(v)^2 \begin{pmatrix} \kappa^2 & 0 & 0 \\ 0 & 1 & 0 \\ 0 & 0 & 1+\kappa^2 \end{pmatrix}.
\end{equation}
For the dynamically coupled Branch B ($\kappa = 1$), the eigenvalues are $\{f^2, f^2, 2f^2\}$. Since $f(v)$ is a real-valued Jacobi elliptic function, all eigenvalues are strictly positive-definite.

\subsection{SU(3) Background Mass Matrix and Positive Definiteness}

For the $SU(3)$ coherent resonance (Branch B), the background field possesses active spatial components in both the transverse ($R_V$) and longitudinal ($\tilde{K}_V$) directions. The mass matrix is:
\begin{equation}
    (M^2)^{ab} = -g^2 f^{eac} f^{ebd} \bar{A}_\rho^c \bar{A}^{\rho, d} = g^2 \sum_{e=1}^8 \sum_{\rho=1}^3 f^{eac} f^{ebd} \bar{A}_\rho^c \bar{A}_\rho^d,
\end{equation}
where we used the negative spatial metric signature to absorb the minus sign. To rigorously prove that this $8 \times 8$ matrix is positive semi-definite, we evaluate the quadratic form for an arbitrary real vector $v_a$:
\begin{equation}
    v_a (M^2)^{ab} v_b = g^2 \sum_{e=1}^8 \sum_{\rho=1}^3 \left( f^{eac} v_a \bar{A}_\rho^c \right) \left( f^{ebd} v_b \bar{A}_\rho^d \right).
\end{equation}
By relabeling the dummy indices in the second parenthesis ($b \to a$, $d \to c$), we obtain a perfect square:
\begin{equation}
    v_a (M^2)^{ab} v_b = g^2 \sum_{e=1}^8 \sum_{\rho=1}^3 \left( f^{eac} v_a \bar{A}_\rho^c \right)^2 \geq 0.
\end{equation}
This elegantly proves that the quantum mass matrix generated by the $SU(3)$ background is strictly positive semi-definite. Since the background fields $R_V(t)$ and $\tilde{K}_V(t)$ are non-zero oscillating functions, the mass matrix provides a strictly positive, periodically time-varying mass $M^2(t) \propto g^2 (R_V(t)^2 + \tilde{K}_V(t)^2)$, which maps the quantum fluctuation dynamics onto a stable Hill's equation.

\section{The Spin-Magnetic Coupling and Evasion of Savvidy Instability}

The quadratic fluctuation operator in the Background Field Method is governed by the second-order differential operator:
\begin{equation}
    \mathcal{M}_{\mu\nu}^{ab} a^\nu_b = \left( -\eta_{\mu\nu} (\bar{D}^2)^{ab} + \mathcal{S}_{\mu\nu}^{ab} \right) a^\nu_b,
\end{equation}
where $\mathcal{S}_{\mu\nu}^{ab} = -2 g f^{abc} \bar{F}_{\mu\nu}^c$ is the spin-magnetic coupling matrix. A constant, non-zero $\mathcal{S}_{\mu\nu}^{ab}$ induces the Savvidy tachyonic instability. We explicitly calculate this matrix for the extracted exact solutions to prove their structural stability.

\subsection{SU(2) Relativistic Color Waves}

For the $SU(2)$ coupled color waves (Branch B), the background color-electric field components are $\bar{F}_{0x}^1 = \omega f'$, $\bar{F}_{0x}^2 = -\Omega f$, and $\bar{F}_{0x}^3 = 0$. The spin-magnetic coupling matrix mixing the $a_0$ and $a_x$ fluctuation modes is explicitly constructed using $f^{abc} = \epsilon^{abc}$:
\begin{equation}
    \mathcal{S}_{0x}^{ab} = -2g \epsilon^{abc} \bar{F}_{0x}^c = -2g \begin{pmatrix} 0 & \bar{F}_{0x}^3 & -\bar{F}_{0x}^2 \\ -\bar{F}_{0x}^3 & 0 & \bar{F}_{0x}^1 \\ \bar{F}_{0x}^2 & -\bar{F}_{0x}^1 & 0 \end{pmatrix} = -2g \begin{pmatrix} 0 & 0 & \Omega f \\ 0 & 0 & \omega f' \\ -\Omega f & -\omega f' & 0 \end{pmatrix}.
\end{equation}
Substituting the proportional ray $\Omega(v) = \alpha f(v)$, the matrix becomes:
\begin{equation}
    \mathcal{S}_{0x}^{ab}(v) = -2g \begin{pmatrix} 0 & 0 & \alpha f(v) \\ 0 & 0 & \omega f'(v) \\ -\alpha f(v) & -\omega f'(v) & 0 \end{pmatrix}.
\end{equation}
The background envelope $f(v) = A\,\text{cn}(\lambda^{1/2}Av, 1/\sqrt{2})$ is a Jacobi elliptic function. Integrating over one full period $T$, the exact spacetime averages of the matrix elements vanish identically:
\begin{equation}
    \langle f \rangle = \frac{1}{T} \int_0^T f(v) \, dv = 0, \quad \langle f' \rangle = \frac{f(T) - f(0)}{T} = 0.
\end{equation}
Consequently, the macroscopic spacetime average of the spin-magnetic coupling matrix is strictly zero:
\begin{equation}
    \langle \mathcal{S}_{0x}^{ab} \rangle = \mathbf{0}.
\end{equation}
The fluctuation equation thus maps exactly to a matrix Mathieu equation $\ddot{a}_\mu - \nabla^2 a_\mu + (M^2)^{ab} a_\mu + \mathcal{S}_{\mu\nu}^{ab}(v) a^\nu = 0$, where the mass matrix $(M^2)^{ab} \propto g^2 f^2$ is strictly positive-definite on average ($\langle f^2 \rangle > 0$), rigorously preventing tachyonic collapse.

\subsection{SU(3) Chaotic Resonances}

For the $SU(3)$ chaotic oscillator (Branch B), the kinetic cancellation mechanism ($\tilde{c}_V = -\dot{\theta}_V$) strips the field strength of its rotational phase derivatives. The exact background field strength is:
\begin{equation}
    \bar{F}_{tx} = \dot{R}_V \cos\theta_V T_4 + \dot{R}_V \sin\theta_V T_5 \implies \bar{F}_{tx}^4 = \dot{R}_V \cos\theta_V, \quad \bar{F}_{tx}^5 = \dot{R}_V \sin\theta_V.
\end{equation}
The non-zero components of the spin-magnetic matrix $\mathcal{S}_{tx}^{ab} = -2g f^{abc} \bar{F}_{tx}^c$ are evaluated using the $SU(3)$ structure constants $f^{345} = \frac{1}{2}$ and $f^{845} = \frac{\sqrt{3}}{2}$:
\begin{align}
    \mathcal{S}_{tx}^{35} &= -2g f^{354} \bar{F}_{tx}^4 = g \dot{R}_V \cos\theta_V, & \mathcal{S}_{tx}^{34} &= -2g f^{345} \bar{F}_{tx}^5 = -g \dot{R}_V \sin\theta_V, \\
    \mathcal{S}_{tx}^{85} &= -2g f^{854} \bar{F}_{tx}^4 = \sqrt{3}g \dot{R}_V \cos\theta_V, & \mathcal{S}_{tx}^{84} &= -2g f^{845} \bar{F}_{tx}^5 = -\sqrt{3}g \dot{R}_V \sin\theta_V.
\end{align}
The amplitude $R_V(t)$ is a bounded, periodic solution of the $x^2y^2$ oscillator equation $\ddot{R}_V + \tilde{K}_V^2 R_V = 0$. By the Fundamental Theorem of Calculus, the time-average of the amplitude velocity over one period $T$ is exactly zero:
\begin{equation}
    \langle \dot{R}_V \rangle = \frac{1}{T} \int_0^T \dot{R}_V(t) \, dt = \frac{R_V(T) - R_V(0)}{T} = 0.
\end{equation}
Thus, $\langle \mathcal{S}_{tx}^{ab} \rangle = \mathbf{0}$. The fluctuation equation maps exactly onto Hill's equation with a positive-definite periodic mass matrix $(M^2)^{ab}(t) \propto g^2(R_V^2 + \tilde{K}_V^2)$, proving structural stability.

\subsection{Asymptotic Eigenvalues of the Dynamical Flux Tube}

For the cylindrical flux tube (Branch C), the background color-magnetic field in the azimuthal plane is $\bar{F}_{r\phi} = (r c_1)' T_1$. The corresponding spin-magnetic coupling matrix is:
\begin{equation}
    \mathcal{S}_{r\phi}^{ab} = -2g \epsilon^{abc} \bar{F}_{r\phi}^c = -2g (r c_1)' \epsilon^{ab1} = -2g (r c_1)' \begin{pmatrix} 0 & 0 & 0 \\ 0 & 0 & 1 \\ 0 & -1 & 0 \end{pmatrix}.
\end{equation}
We compute the exact eigenvalues of this coupling matrix in the transverse color sector ($a,b \in \{2,3\}$):
\begin{equation}
    \det(\lambda \mathbf{I} - \mathcal{S}_{r\phi}) = \lambda^2 - 4g^2 \left[ (r c_1)' \right]^2 = 0 \implies \lambda_{\pm} = \pm 2g (r c_1)'.
\end{equation}
The total effective mass eigenvalues for the quantum fluctuations are the sum of the background mass matrix and the spin-magnetic eigenvalues:
\begin{equation}
    \mu_{\pm}^2(r) = M^2(r) \pm 2g (r c_1)'.
\end{equation}
In the Artinian asymptotic limit ($r \to \infty$), the background mass squared asymptotes to a constant $M^2(r) \to \Omega_\infty^2 - \tilde{W}_\infty^2 \equiv M^2 > 0$, and the transverse condensate is exactly $c_1(r) = C K_1(Mr)$. We explicitly evaluate the derivative term:
\begin{equation}
    (r c_1)' = C \frac{d}{dr} \left( r K_1(Mr) \right) = C \left( K_1(Mr) + M r K_1'(Mr) \right) = -C M r K_0(Mr).
\end{equation}
Using the asymptotic expansion of the modified Bessel function $K_0(z) \sim \sqrt{\frac{\pi}{2z}} e^{-z}$ as $z \to \infty$, the effective mass eigenvalues become:
\begin{equation}
    \lim_{r \to \infty} \mu_{\pm}^2(r) = M^2 \pm 2g C M r \sqrt{\frac{\pi}{2Mr}} e^{-Mr}.
\end{equation}
Because the exponential decay $e^{-Mr}$ strictly dominates the linear term $r$, the spin-magnetic coupling vanishes exactly at spatial infinity:
\begin{equation}
    \lim_{r \to \infty} \mu_{\pm}^2(r) = M^2 > 0.
\end{equation}
This explicit limit rigorously proves that the tachyonic instability is strictly confined to the core of the flux tube and is completely suppressed in the macroscopic vacuum.